\documentclass[longauth]{aa}  

\usepackage[varg]{txfonts}
\usepackage{graphicx}
\usepackage{amsmath}
\usepackage{amssymb}
\usepackage{color}
\usepackage{xcolor}
\usepackage{natbib}
\bibpunct{(}{)}{;}{a}{}{,} 
\usepackage{url}
\usepackage[color=blue!20, bordercolor=orange, disable]{todonotes}
\usepackage{multirow}
\usepackage{threeparttable}      
\usepackage{booktabs} 
\usepackage{soul}
\usepackage{float}
\usepackage[colorlinks=true]{hyperref}
\usepackage{tablefootnote}
\hypersetup{colorlinks=true,citecolor=blue}

\begin{document} 
   \title{NIRPS detection of delayed atmospheric escape from the warm and misaligned Saturn-mass exoplanet WASP-69\,b\thanks{Based in part on Guaranteed Time Observations collected at the European Southern Observatory under ESO program 111.2506 by the NIRPS Consortium.}}

\author{
Romain Allart\inst{1,*,\thanks{SNSF Postdoctoral Fellow}},
Yann Carteret\inst{2},
Vincent Bourrier\inst{2},
Lucile Mignon\inst{3,2},
Fr\'ed\'erique Baron\inst{1,4},
Charles Cadieux\inst{1},
Andres Carmona\inst{3},
Christophe Lovis\inst{2},
Hritam Chakraborty\inst{2},
Elisa Delgado-Mena\inst{5,6}
% ...
}

\institute{
\inst{1}Institut Trottier de recherche sur les exoplan\`etes, D\'epartement de Physique, Universit\'e de Montr\'eal, Montr\'eal, Qu\'ebec, Canada\\
\inst{2}Observatoire de Gen\`eve, D\'epartement d’Astronomie, Universit\'e de Gen\`eve, Chemin Pegasi 51, 1290 Versoix, Switzerland\\
\inst{3}Univ. Grenoble Alpes, CNRS, IPAG, F-38000 Grenoble, France\\
\inst{4}Observatoire du Mont-M\'egantic, Qu\'ebec, Canada\\
\inst{5}Centro de Astrobiolog\'ia (CAB), CSIC-INTA, Camino Bajo del Castillo s/n, 28692, Villanueva de la Ca\~nada (Madrid), Spain\\
\inst{6}Instituto de Astrof\'isica e Ci\^encias do Espa\c{c}o, Universidade do Porto, CAUP, Rua das Estrelas, 4150-762 Porto, Portugal\\
\inst{7}Departamento de F\'isica e Astronomia, Faculdade de Ci\^encias, Universidade do Porto, Rua do Campo Alegre, 4169-007 Porto, Portugal\\
\inst{8}Department of Physics, University of Toronto, Toronto, ON M5S 3H4, Canada\\
\inst{9}Departamento de F\'isica Te\'orica e Experimental, Universidade Federal do Rio Grande do Norte, Campus Universit\'ario, Natal, RN, 59072-970, Brazil\\
\inst{10}Department of Physics \& Astronomy, McMaster University, 1280 Main St W, Hamilton, ON, L8S 4L8, Canada\\
\inst{11}Department of Physics, McGill University, 3600 rue University, Montr\'eal, QC, H3A 2T8, Canada\\
\inst{12}Department of Earth \& Planetary Sciences, McGill University, 3450 rue University, Montr\'eal, QC, H3A 0E8, Canada\\
\inst{13}Centre Vie dans l’Univers, Facult\'e des sciences de l’Universit\'e de Gen\`eve, Quai Ernest-Ansermet 30, 1205 Geneva, Switzerland\\
\inst{14}Instituto de Astrof\'isica de Canarias (IAC), Calle V\'ia L\'actea s/n, 38205 La Laguna, Tenerife, Spain\\
\inst{15}Departamento de Astrof\'isica, Universidad de La Laguna (ULL), 38206 La Laguna, Tenerife, Spain\\
\inst{16}European Southern Observatory (ESO), Karl-Schwarzschild-Str. 2, 85748 Garching bei M\"unchen, Germany\\
\inst{17}Space Research and Planetary Sciences, Physics Institute, University of Bern, Gesellschaftsstrasse 6, 3012 Bern, Switzerland\\
\inst{18}Consejo Superior de Investigaciones Cient\'ificas (CSIC), E-28006 Madrid, Spain\\
\inst{19}Bishop's Univeristy, Dept of Physics and Astronomy, Johnson-104E, 2600 College Street, Sherbrooke, QC, Canada, J1M 1Z7\\
\inst{20}Department of Physics and Space Science, Royal Military College of Canada, PO Box 17000, Station Forces, Kingston, ON, Canada\\
\inst{21}Instituto de Astrof\'isica e Ci\^encias do Espa\c{c}o, Faculdade de Ci\^encias da Universidade de Lisboa, Campo Grande, 1749-016 Lisboa, Portugal\\
\inst{22}Departamento de F\'isica da Faculdade de Ci\^encias da Universidade de Lisboa, Edif\'icio C8, 1749-016 Lisboa, Portugal\\
\inst{23}Centre of Optics, Photonics and Lasers, Universit\'e Laval, Qu\'ebec, Canada\\
\inst{24}Center for Space and Habitability, University of Bern, Gesellschaftsstrasse 6, 3012 Bern, Switzerland\\
\inst{25}Aix Marseille Univ, CNRS, CNES, LAM, Marseille, France\\
\inst{26}Departamento de F\'isica, Universidade Federal do Cear\'a, Caixa Postal 6030, Campus do Pici, Fortaleza, Brazil\\
\inst{27}European Southern Observatory (ESO), Av. Alonso de Cordova 3107, Casilla 19001, Santiago de Chile, Chile\\
\inst{28}Plan\'etarium de Montr\'eal, Espace pour la Vie, 4801 av. Pierre-de Coubertin, Montr\'eal, Qu\'ebec, Canada\\
\inst{29}Lund Observatory, Division of Astrophysics, Department of Physics, Lund University, Box 118, 221 00 Lund, Sweden\\
\inst{30}York University, 4700 Keele St, North York, ON M3J 1P3\\
\inst{31}Herzberg Astronomy and Astrophysics Research Centre, National Research Council of Canada\\
\inst{32}University of British Columbia, 2329 West Mall, Vancouver, BC, canada, v6t 1z4\\
\inst{33}Western University, Department of Physics \& Astronomy and Institute for Earth and Space Exploration, 1151 Richmond Street, London, ON N6A 3K7, Canada\\
\inst{34}Light Bridges S.L., Observatorio del Teide, Carretera del Observatorio, s/n Guimar, 38500, Tenerife, Canarias, Spain\\
\inst{35}University Observatory, Faculty of Physics, Ludwig-Maximilians-Universit\"at M\"unchen, Scheinerstr. 1, 81679 Munich, Germany\\
\inst{36}Institute of Space Sciences (ICE, CSIC), Carrer de Can Magrans S/N, Campus UAB, Cerdanyola del Valles, E-08193, Spain\\
\inst{37}Institut d’Estudis Espacials de Catalunya (IEEC), 08860 Castelldefels (Barcelona), Spain\\
\inst{38}Laboratoire Lagrange, Observatoire de la C\^ote d’Azur, CNRS, Universit\'e C\^ote d’Azur, Nice, France\\
\inst{*}\email{romain.allart@umontreal.ca}
}

   \date{Received October 8, 2024; accepted December 17, 2024}

 
  \abstract
   {Near-infrared high-resolution \'echelle spectrographs unlock access to fundamental properties of exoplanets, from their atmospheric escape and composition to their orbital architecture, which can all be studied simultaneously from transit observations.}
   {We present the first results of the newly commissioned ESO near-infrared spectrograph, \textit{Near-InfraRed Planet Searcher} (NIRPS), from three transits of the well-studied warm Saturn WASP-69\,b. Our goals are to measure the orbital architecture of the planet through the Rossiter-McLaughlin (RM) effect and its atmospheric escape through the 1083\,nm helium triplet.}
   {We used the RM Revolutions technique to better constrain the orbital architecture of the system. We extracted the high-resolution helium absorption profile to study its spectral shape and temporal variations.  Then, we made 3D simulations from the EVE code to fit the helium absorption time series.}
   {We measure a slightly misaligned orbit for WASP-69\,b (3D spin-orbit angle of 28.7$^{+6.1}_{-5.3}$ $^{\circ}$). We confirm the detection of helium with an average excess absorption of 3.17$\pm$0.05\% (maximum of 4.02\%). The helium absorption is spectrally and temporally resolved, extends to high altitudes and has a strong velocity shift up to $-$29.5$\pm$2.5\,km$\cdot$s$^{-1}$ 50 minutes after egress. The signature cannot be explained by a thermosphere alone and thus requires 3D modeling of the thermosphere and exosphere. EVE simulations put constraints on the mass loss of $2.25\cdot 10^{11}$~g$\cdot$s$^{-1}$ and hint at reactive chemistry within the cometary-like tail and interaction with the stellar winds that allow the metastable helium to survive longer than expected. }
   {Our results suggest that WASP-69\,b is going through a transformative phase of its history by losing mass while evolving on a misaligned orbit, similar to a growing number of Neptunian worlds. This work shows how combining multiple observational tracers such as orbital architecture, atmospheric escape, and composition is critical to understand exoplanet demographics and their formation and evolution. We demonstrate that NIRPS in the near-infrared can reach precisions similar to HARPS in the optical for RM studies, and the high data quality of NIRPS leads to unprecedented atmospheric characterization. Therefore, the addition of NIRPS to HARPS on the ESO 3.6m makes it the driving force of such new studies. The high stability of NIRPS combined with the  large Guaranteed Time Observation (GTO) available for its consortium enables in-depth studies of exoplanets as well as large population surveys.}

   \keywords{Planetary systems -- Planets and satellites: atmospheres, gaseous planets, individual: WASP-69b -- Instrumentation: spectrographs -- Methods: observational -- Techniques: spectroscopic}
   \titlerunning{WASP-69b seen by NIRPS}
   \authorrunning{R. Allart, Y. Carteret, V. Bourrier et al. }
   \maketitle
%

\section{Introduction}\label{sec:intro}
\nolinenumbers

The exoplanet population is sculpted by how planets form and evolve. The Neptunian desert \citep{lecavelier_des_etangs_diagram_2007,mazeh_dearth_2016} and savanna \citep{bourrier_dream_2023} are unbiased observational evidence of a dearth of hot Neptune-mass planets at close-by separation from their host stars, which extends to lower temperatures with a sparse population of warm Neptunes. \cite{helled_mass_2023} and \cite{venturini_formation_2017} define Neptunian planets as the population of gas giants that never reached the runaway gas accretion phase during their formation. This led to their higher metallicity and broader compositional diversity \citep{moses_compositional_2013}. Therefore, Neptunian planets have masses ranging from $\sim$12\,M$_{\oplus}$ (5\,M$_{\oplus}$ below Neptune's mass) to $\sim$100\,M$_{\oplus}$ (Saturn's mass). Due to their specific mass regime, these planets are more likely than Jupiter-mass planets to lose their atmosphere under the strong X-ray and Extreme Ultraviolet (XUV) irradiation from their host star during their lifetime \citep[e.g.,][]{ehrenreich_giant_2015,bourrier_hubble_2018,allart_spectrally_2018,allart_high-resolution_2019} and could even become bare surfaces. In addition to having massive atmospheric outflows and a broad diversity of atmospheric composition, observations show that Neptunian worlds are more likely to undergo high eccentricity migration \citep{castro_ridge_2024}. Several have been observed to lose their atmosphere while having misaligned (up to pole-on or retrograde), and often eccentric, orbits \citep[e.g.,][]{bourrier_orbital_2018,Rubenzahl_retro_2021,stefansson_polar_2022,bourrier_dream_2023}. This diversity of orbital architectures, atmospheric escapes, and compositions makes Neptunian worlds a unique population to understand how exoplanets form and evolve.\\
One of the best ways to study exoplanet evaporation is through the near-infrared metastable helium triplet \citep{Seager_transmission_2000,oklopcic_new_2018}, which is not absorbed by the interstellar medium \citep{Indriolo_ISM_2009} and provides unique access to the exosphere, the atmospheric layer no longer gravitationally bound to the exoplanets and where mass loss occurs. Ground-based near-infrared high-resolution spectrographs (e.g., CARMENES, GIANO, NIRSPEC, or SPIRou) have led to several unamiguous spectrally and temporally resolved detections \citep[e.g.,][]{allart_spectrally_2018,allart_high-resolution_2019,allart_homogeneous_2023,nortmann_ground-based_2018,Kasper_Helium_2020,Kirk_Keck_2022,Zhang_mini_2023,Zhang_Giant_2023,Orell-Miquel_MOPYS_2024,guilluy_gaps_2024}, highlighting the use of the helium triplet as a robust atmospheric tracer.\\

The \textit{Near-InfraRed Planet Searcher} (NIRPS) consortium \citep{Bouchy_2025}, in exchange for building the instrument and coordinating the schedule and observations of the European Southern Observatory (ESO) 3.6\,m telescope, was granted 725 nights of guaranteed time observations (GTO) over five years by ESO. The consortium, led by Canada and Switzerland, with contributions from France, Portugal, Spain, and Brazil, allocated 225 nights for the in-depth characterization of exoplanets. Its Work-Package three (WP3) aims at collecting high-fidelity, high-signal-to-noise transmission and emission spectra. In addition, large comprehensive atmospheric and orbital architecture surveys are planned. Combined, more than 75 exoplanets from ultra-hot Jupiters to temperate terrestrial planets will be observed, with a prominent fraction of time spent on gas giant planets, from Neptunian worlds to Jupiter-mass planets. For each planet, the goals are to measure its orbital architecture, atmospheric escape, dynamics, and composition. Once brought to the population level, these constraints will provide critical trends to refine formation and evolution models. During NIRPS's first six months of operation, its consortium observed three transits of WASP-69\,b, a well-known warm Saturn-mass planet, as part of its transit survey. This paper presents the first constraints on the orbital architecture and atmospheric escape obtained with NIRPS simultaneously to the \textit{High Accuracy Radial velocity Planet Searcher} (HARPS). A subsequent paper will focus on the atmospheric composition once more data are collected.\\

WASP-69\,b is a warm Saturn-mass planet orbiting an active K dwarf \citep{anderson_three_2014}, and Table\,\ref{table:params} summarizes the properties of the system. \cite{casasayas-barris_detection_2017} inferred the orbital architecture of the system through the Rossiter-McLaughlin (RM) effect and concluded that it is an aligned system ($\lambda$ = 0.4$^{+2.0}_{-1.9}$ $^{\circ}$). WASP-69b is one of the two first exoplanets (alongside HAT-P-11\,b, \citealt{allart_spectrally_2018}) to have a measured excess absorption of helium obtained at high resolution with CARMENES \citep{nortmann_ground-based_2018}. The authors measured a clear maximum excess absorption of 3.59$\pm$0.19\%, detected in two independent visits at 3.96$\pm$0.25\% and 3.00$\pm$0.31\%. The signature is slightly blueshifted ($-3.58\pm$0.23\,km$\cdot$s$^{-1}$) over the whole transit, but shifts from 1.4$\pm$0.4\,km$\cdot$s$^{-1}$ during ingress to $-$10.7$\pm$1.1\,km$\cdot$s$^{-1}$ during the post-transit absorption, which lasted for 22 minutes after egress (corresponding to a sky-projected extension of the tail of up to 2.2\,R$_p$.). The in-transit excess absorption was later confirmed at low spectral resolution by \cite{vissapragada_constraints_2020} and \cite{levine_exoplanet_2024}, and also observed at high spectral resolution with CFHT/SPIRou \citep{allart_homogeneous_2023,masson_probing_2024}, Keck/NIRSPEC \citep{tyler_wasp-69bs_2024} and TNG/GIANO-B \citep{guilluy_gaps_2024}. The SPIRou transit \citep{allart_homogeneous_2023} revealed an averaged absorption of 2.21$\pm$0.25\,\% with a maximum at $\sim$3.41\,\% but no clear post-transit absorption was detected, probably due to strong systematics and the shorter post-transit baseline. The helium signature detected in the single Keck/NIRSPEC transit \citep{tyler_wasp-69bs_2024} has an average absorption of 2.7$\pm$0.4\% and a blueshift of $-$5.9$\pm$1.0\,km$\cdot$s$^{-1}$. They also measured post-transit absorption over their complete post-transit sequence (1.28h). Assuming that absorption from the tail lasts for the full 1.28\,h, the tail would extend by 7\,R$_p$ in sky projection, far beyond the planet's Roche lobe ($\sim$2.7R$_p$). In addition, their post-transit signal has a net blueshift of $-23.3\pm0.9$\,km$\cdot$s$^{-1}$. \cite{guilluy_gaps_2024} analyzed three GIANO-B transits, estimated at 3.91$\pm$0.22\% the helium signature contrast, and confirmed post-transit absorption lasting for $\sim$50 minutes. Moreover, they show evidence for variable signal strength from transit to transit (similar to \citealp{nortmann_ground-based_2018}) which correlates with an emission signal in the H$\alpha$ line. Both \cite{allart_homogeneous_2023} and \cite{guilluy_gaps_2024} concluded that stellar variability alone cannot account for the observed helium signature or the post-transit absorption.\\
The most recent study of the WASP-69\,b helium triplet was conducted by \cite{levine_exoplanet_2024} with an ultra-narrow filter on Palomar/WIRC. The authors measured a weaker helium absorption than all previous published transits at both low and high spectral resolution. Coincidentally, they measured this lower absorption signal on the transit occurring on the night of 24 August 2023, which we also observed with NIRPS and which we analyze here.\\ 
Finally, 1D and 3D modeling estimated mass-loss rates ranging from 0.2\,M$_{\oplus}\cdot$Gyr$^{-1}$ to 1.0\,M$_{\oplus}\cdot$Gyr$^{-1}$ \citep{vissapragada_constraints_2020,wang_metastable_2021,lampon_characterisation_2023,tyler_wasp-69bs_2024} depending on the datasets and model frameworks used. However, as the models used in the previous studies were modelling the thermosphere of the exoplanet and not its exosphere, the post-transit signal could not be adequately reproduced. \\

Therefore, WASP-69\,b is a prime target for the NIRPS consortium to validate the instrument performances and deepen our understanding of this keystone system. We describe the NIRPS, HARPS, and HARPS-N observations in Section\,\ref{sec:obs}. Section\,\ref{sec:stellar} presents updated stellar parameters, while Section\,\ref{sec:photometry} describes the simultaneous photometry with ExTrA and EulerCam along with updated system parameters. Section\,\ref{sec:orb_arch} analyzes the orbital architecture of the WASP-69 system, while the helium analysis and modeling are detailed in Section\,\ref{sec:helium}. Section\,\ref{sec:discussion} situates WASP-69\,b in the context of the known population of exoplanets with escaping helium particles and misaligned orbits. We conclude in Section\,\ref{sec:conclusion}.

\begin{table*}[t]
\begin{minipage}[tbh!]{\textwidth}
\caption{Parameters of the WASP-69 system}
\label{table:params}
\small
\centering
\begin{tabular}{l l c l}
\toprule \toprule
Parameter & Symbol  [Unit] & Value & Reference \\
\toprule
\toprule
\multicolumn{4}{l}{\textit{Stellar Parameters}} \\
Apparent magnitude & $\mathrm{m_v}$  & 9.87 & \cite{hog_tycho_2000}\\
Luminosity & $\mathrm{L_\star}$ [$\mathrm{L_\odot}$] & 0.33 & This work\\
Distance & d [pc] & 50.3 & \cite{Sousa-2021} \\
Stellar age & Age [$\mathrm{Gyr}$] & 1 -- 2.6 & This work\\
Stellar mass & $\mathrm{M_\star}$ [$\mathrm{M_\odot}$] & 0.83 $\pm$ 0.05 & This work\\
Stellar radius & $\mathrm{R_\star}$ [$\mathrm{R_\odot}$] & 0.801 $\pm$ 0.015 & This work\\ 
Effective temperature & $\mathrm{T_{eff}}$ [K] & 4792 $\pm$ 158 & This work\\
Metallicity & [Fe/H] [dex] & 0.35 $\pm$ 0.07 & This work \\
Turbulent velocity & $\mathrm{v_{turb}}$ [km$\cdot$ s$^{-1}$] &0.55 $\pm$ 0.008 & This work\\
Surface gravity$^{\textcolor{red}{1}}$ & $\mathrm{log\,g_\star}$ [cgs] &  4.16 $\pm$ 0.037 & This work\\
Surface gravity$^{\textcolor{red}{2}}$ & $\mathrm{log\,g_\star}$ [cgs] &  4.57 $\pm$ 0.037 & This work\\
Activity indicator & $\log \mathrm{R'_{HK}}$ [dex] & -4.597 $\pm$ 0.001 & This work\\
Limb-darkening coefficient & $u_\mathrm{1}$ & 0.2669 &  This work\\
Limb-darkening coefficient & $u_\mathrm{2}$ & 0.2657 &  This work\\
Spin inclination & $\mathrm{i_\star}$ [deg] & 60.8$^{+5.0}_{-6.1}$   & This work \\
Equatorial rotation period & $\mathrm{P_{eq}}$ [days] & 23.07 $\pm$ 0.16  & \cite{anderson_three_2014} \\
Projected rotational velocity & $\mathrm{v sin i_\star}$ [km$\cdot$ s$^{-1}$]& 1.54$\pm$0.05 & This work\\
Stellar RV semi-amplitude & $\mathrm{K_\star}$ [m$\cdot$ s$^{-1}$] & 38.1 $\pm$ 2.4  & \cite{anderson_three_2014}\\
\toprule
\multicolumn{4}{l}{\textit{Planetary Parameters}} \\
Planet mass & $\mathrm{M_p}$ [$\mathrm{M_\oplus}$] & 82.58 $\pm$ 5.37 & This work \\
Planet mass & $\mathrm{M_p}$ [$\mathrm{M_{Jup}}$] & 0.26 $\pm$ 0.02 & This work \\
Planet radius & $\mathrm{R_p}$ [$\mathrm{R_\oplus}$] & 11.21 $\pm$ 0.22 & This work\\
Planet radius & $\mathrm{R_p}$ [$\mathrm{R_{Jup}}$] & 1.00 $\pm$ 0.02 & This work\\
Equilibrium temperature$^{\textcolor{red}{3}}$ & $\mathrm{T_{eq}}$ [K] & 971 $\pm$ 33 & This work \\
Projected spin-orbit angle & $\lambda$ [deg] & 0.05 $\pm$ 1.10 &  This work\\
3D spin-orbit angle$^{\textcolor{red}{4}}$ & $\psi$ [deg]  &  29.2$^{+6.1}_{-5.0}$   & This work \\
Mid-transit time & $\mathrm{T_c}$ [$\mathrm{BJD_{TDB}}$]& 2455748.83428 $\pm$ 0.00013 & This work \\
Orbital period & P [d] & 3.86813881 $\pm$ 0.00000013 & This work   \\
Transit duration (t$_1$ to t$_4$) & $\mathrm{t_{14}}$ [h] & 2.1869 $\pm$ 0.0053 & This work \\
Orbital semi-major axis & a [$R_\mathrm{\star}$] & 12.17 $\pm$ 0.07 & This work \\
Orbital semi-major axis & a [AU] & 0.0453 $\pm$ 0.0008 & This work \\
Orbit inclination & $\mathrm{i_p}$ [deg] & 86.79 $\pm$ 0.04 & This work \\
Eccentricity & e & 0 (fixed) & This work\\
Periastron argument & $\omega$ & 90 (fixed) & This work\\

\bottomrule
\end{tabular}
\tablefoot{
$^{\textcolor{red}{1}}$: surface gravity from spectroscopy;
$^{\textcolor{red}{2}}$: surface gravity calibrated with \cite{Mortier-2014} and adopted in the this study;
$^{\textcolor{red}{3}}$: assuming a null albedo;
$^{\textcolor{red}{4}}$: The 3D spin–orbit angle is combined over the two degenerate configurations for the stellar inclination (see text).
}
\end{minipage}
\end{table*}

\section{Observations}\label{sec:obs}
In this paper, we used NIRPS and HARPS simultaneous observations obtained by the NIRPS consortium, which have been combined with HARPS-N and HARPS archival data to study the Rossiter-McLaughlin effect (see section\,\ref{sec:orb_arch}). The NIRPS data are, in addition, used to study the near-infrared helium triplet (Section \ref{sec:helium}). We describe hereafter these observations.

\subsection{Simultaneous NIRPS and HARPS observations}

\begin{figure}
\includegraphics[width=\columnwidth]{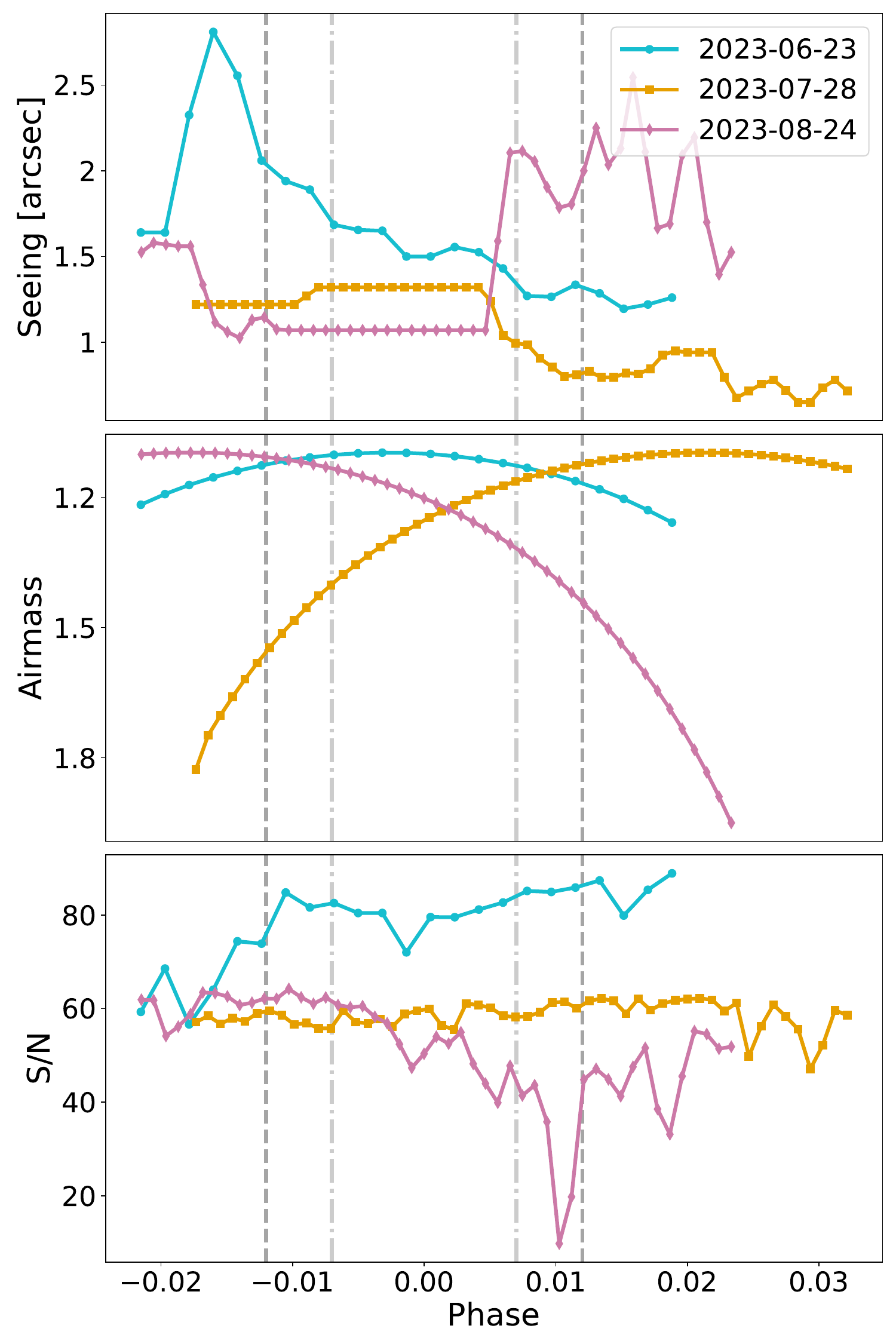}
\centering
\caption[]{Observing conditions during the three transits of WASP-69\,b observed with NIRPS and HARPS simultaneously. From \textit{top} to \textit{bottom}: Dimm seeing, airmass, and S/N of order 14 of NIRPS, where the helium triplet is located. Recording of the seeing during the transits partly ceased to function, leading to constant values. Grey vertical lines correspond to the transit contact points t$_1$, t$_4$ (dashed), t$_2$, and t$_3$ (dot-dashed).}
\label{fig:condtions}
\end{figure}

The NIRPS consortium observed three transits of WASP-69\,b simultaneously with NIRPS \citep{bouchy_near-infrared_2017,artigauspie,Bouchy_2025} and HARPS (\citealp{mayor_setting_2003}), both installed on the 3.6\,m ESO telescope at La Silla, Chile. NIRPS is a near-infrared (0.98--1.8\,$\mu$m), fiber-fed, stabilized, high-resolution \'echelle spectrograph assisted by adaptive optics (AO). HARPS is an optical (380--690\,nm), fiber-fed, stabilized, high-resolution \'echelle spectrograph. The NIRPS GTO consortium collected the observations (Prog: 111.2506.001, PI: F. Bouchy) in the high-efficiency (HE) mode ($\mathcal{R}\sim$75200, 0.9" fiber size on sky) and high-accuracy mode (HAM) for NIRPS and HARPS, respectively. For NIRPS, the high-efficiency mode was selected over its high-accuracy mode as it reduces the impact of modal noise at the reddest wavelengths, while the loss in spectral resolution is minimal \citep{artigauspie,Bouchy_2025}. Fibers B of both instruments were used to monitor the sky. We observed WASP-69 during the following nights\footnote{A fourth attempt was performed on the night of 2023-07-24 but had to be interrupted due to bad weather. The data quality is insufficient to use the dataset.}: 2023-06-23, 2023-07-28, and 2023-08-24. Table \ref{observation W69} and Fig.\,\ref{fig:condtions} summarize the observing conditions on those nights. ESO period 111 suffered from overall worse weather than previous years, which translates into a large seeing values, but the NIRPS AO system limited the loss of signal-to-noise ratio (S/N). Nonetheless, the first transit (2023-06-23) was observed with a larger exposure time due to large seeing values to reach higher S/N. In addition, two spectra obtained during the transit egress on 2023-08-24 have lower than expected S/N because the AO loop opened due to increasing seeing. Overall, the data quality is excellent and in agreement with previous observations (see Section\,\ref{sec:discussion}) despite the high seeing \citep{artigauspie,Bouchy_2025}.\\

\begin{table}[h]
\centering
\caption{Summary of the NIRPS observations.}
\begin{tabular}{lccc}
\toprule \toprule
Observing nights & 2023-06-23 & 2023-07-28 & 2023-08-24\\
\toprule \toprule
Total spectra                                                                                                               & 23            & 54            & 49  \\
N$\mathrm{_{In}}$ spectra                                                                                                          & 13            & 26            & 25   \\
N$\mathrm{_{Out}}$ spectra$^\dagger$    & 10 (4)        & 28 (14)       & 24 (13)  \\
 t$\mathrm{_{exp}}$ [s]                                                                                                     & 600           & 300           & 300  \\
 Airmass                                                                                                                    & 1.1 -- 1.3    & 1.1 --1.8     & 1.1 -- 2.0  \\
 Seeing [arcsec]                                                                                                                     & 1.2 -- 2.8    & 0.6 -- 1.3    & 1.0 -- 2.5 \\
 S/N@1083nm                                                                                                                 & 57 -- 89      & 47 -- 62      & 10 -- 64  \\
\hline
\end{tabular}
\tablefoot{
$^\dagger$: Total number of spectra and number of spectra impacted by post-transit absorption.}
\label{observation W69}
\end{table}

\subsection{NIRPS data reduction}

The NIRPS data were reduced using both version 3.2.0 of the NIRPS Data Reduction Software (DRS) pipeline adapted from the ESPRESSO pipeline \citep{pepe_espresso_2021} and version 0.7.288 of APERO adapted from SPIRou \citep{cook_apero_2022}. As discussed in \cite{Bouchy_2025}, both pipelines perform calibrations and pre-processing to remove detector effects, including dark, bad pixel, and background, and perform detector non-linearity corrections \citep{artigau_h4rg_2018}, localization of the orders, geometric changes in the image plane, correction of the flat, hot pixel, cosmic ray correction, and wavelength calibration (using both a hollow-cathode UNe lamp and the Fabry-P\'erot \'etalon; \citealt{hobson_spirou_2021}). This is done using a combination of daily and reference calibrations. The result is an optimally extracted spectrum referred to as extracted 2D spectra: \texttt{S2D} and \texttt{e2ds}, respectively. Both pipelines produce compatible 2D extracted spectra but with slightly different dimensions: NIRPS-DRS extracts \texttt{S2D} spectra of 71 orders by 4084 pixels, while APERO extracts \texttt{e2ds} spectra of 75 orders by 4096 pixels. Once the 2D spectra are extracted, both pipelines slightly differ in their post-processing with independent telluric corrections. The NIRPS DRS pipeline corrects telluric absorption lines using the model\footnote{https://github.com/RomainAllart/Telluric\_correction} described in \cite{allart_automatic_2022} and emission lines are corrected using a high-S/N master spectrum comprised of the locations of only the emission features. Each emission line from the master spectrum is then scaled accordingly and subtracted locally from the science spectrum of the observed star. The APERO pipeline telluric correction consists of a two-step process. We first adjust a telluric absorption model using the TAPAS atmosphere model \citep{bertaux_TAPAS_2014}. This adjustment is performed by allowing only two degrees of freedom, one for water opacity and one for all the ``dry'' atmospheric components (O$_3$, O$_2$, CO$_2$ N$_2$O, and CH$_4$) at their nominal fractional atmospheric abundances. A second-order correction is applied with residuals from the first correction as derived from regular observations of telluric standards. These residuals correlate strongly with the water and ``dry'' coefficients and arise from inaccuracies in the atmosphere models, such as line strengths and broadening profiles, and can be applied blindly rather than adjusted once the first-order correction coefficients are known. In Appendix\,\ref{app:pipelines_comp}, we demonstrate that both pipelines produce compatible transmission spectra of WASP-69\,b around the helium triplet. 

\subsection{Archival HARPS and HARPS-N observations}
In addition to the three HARPS transits obtained simultaneously to NIRPS, we used one archival HARPS transit observed on 2012-06-21 (ProgID: 089.C-0151(B), PI: Triaud), and two archival HARPS-N transits observed on 2016-06-03 and 2016-08-04 (\citealp{casasayas-barris_detection_2017}). Spectra were extracted from the detector images, corrected, and calibrated using versions 3.0.1 of HARPS-N and 3.0.0  of HARPS of the Data Reduction Software (DRS) pipelines adapted from the ESPRESSO pipeline (\citealt{pepe_espresso_2021}).

\section{Stellar parameters}\label{sec:stellar}

To derive the spectroscopic stellar parameters ($\mathrm{T_{eff}}$, $\mathrm{log\,g_\star}$, microturbulence, [Fe/H]) we first co-added all the individual HARPS exposures to create a combined spectrum with higher signal-to-noise ratio. We used ATLAS model atmospheres \citep{kurucz} and followed the ARES+MOOG methodology as described in \cite{Sousa-2011}, \cite{Santos-2013}, and \cite{Sousa-2021}. In short, the parameters were measured from the equivalent widths of a set of 71 \ion{Fe}{i} and 9 \ion{Fe}{ii} lines, assuming Local Thermodynamic Equilibrium (LTE) and excitation and ionization equilibrium. The line equivalent widths were measured using the ARES code \citep[][]{Sousa-2011}. Given the temperature of the star, we used the line-list of \citet[][]{Tsantaki-2013}, optimized for the derivation of stellar parameters for K dwarfs. We point to the abovementioned papers for more details, including the derivation of the uncertainties. The value for the surface gravity was also corrected from systematic effects \citep[see][]{Mortier-2014}. Both the spectroscopic value and the corrected one are listed in Table\,\ref{table:params}. The obtained parameters are compatible with literature values listed in the SWEET-Cat database\footnote{https://sweetcat.iastro.pt}. From the parameters obtained, we also estimated the stellar mass (0.81$\pm$0.08\,$\mathrm{M_\odot}$) using the calibrations presented in \citet[][]{Torres-2010}. The derived temperature was further used to estimate the radius of the star (0.83\,$\mathrm{R_\odot}$) from basic principles, using also the luminosity derived from the visual magnitude of the star and its parallax from \textit{Gaia} DR3 \citep[][]{GaiaDR3}, and the bolometric correction from \citet[][]{Flower-1996}. 

Furthermore, we estimated the mass and radius with the tool PARAM1.3\footnote{\url{https://stev.oapd.inaf.it/cgi-bin/param}} which uses PARSEC isochrones \citep{bressan12}, the magnitude \textit{V}, parallax from \textit{Gaia} DR3 \citep{GaiaDR3} and the spectroscopic T$_{\mathrm{eff}}$ and [Fe/H]. We obtain a mass M=0.831 $\pm$ 0.020\,$\mathrm{M_\odot}$ and a radius R=0.771 $\pm$ 0.017\,$\mathrm{R_\odot}$. Based on \cite{tayar_2022}, we added a systematic error of 5\% on the stellar mass leading to 0.831 $\pm$ 0.045\,$\mathrm{M_\odot}$. This mass is in agreement (within errors) with the abovementioned mass, but the radius is much smaller. Therefore, we derived the stellar radius to be 0.801$\pm$0.015\,$\mathrm{R_\odot}$ from the stellar density measured on the light curve (see Section\,\ref{subsec:sys_params}). On the other hand, the age provided by this tool is quite uncertain (4.675 $\pm$ 4.012 Gyr, as expected for K dwarfs using isochrones) and a visual inspection of the spectrum revealed emission in the core of the $\ion{Ca}{ii}$ H$\&$K lines, which could be indicative of a younger age. Therefore, we explored other ways of estimating the age.

To further assess the age of the star, we derived the Li abundance by performing spectral synthesis around the Li doublet region using the same model atmospheres and radiative transfer code to obtain the spectroscopic parameters \citep[see][for further details]{delgadomena14}. We can only determine an upper limit for the Li abundance: A(Li)\footnote{${\rm A(Li)}=\log[N({\rm Li})/N({\rm H})]+12$}\,<\,0.2\,dex, which lays at the upper envelope of Li abundances for main sequence stars of similar effective temperature. This might indicate a young age, given that cool stars have thick convective envelopes that quickly burn the Li by internal mixing. By comparing this Li abundance with observations of stars with similar $\mathrm{T_{eff}}$ within open clusters, we can set a lower age limit of 650 Myr \citep[e.g., Praesepe in ][]{cummings17}.

Using the same co-added spectrum of all observations from HARPS, we estimated the average activity level to be $\log R'_\mathrm{HK} = -4.597 \pm 0.001$ dex, using \verb+ACTIN2+\footnote{\url{https://github.com/gomesdasilva/ACTIN2}} \citep{gomesdasilva2018, gomesdasilva2021} and the methodology described in \citet{gomesdasilva2021}.
This activity level agrees with that obtained in \citet{anderson_three_2014} of $-4.54$ dex and suggests an active star. The chromospheric age was estimated via the activity-mass-metallicity-age relation of \citet{lorenzo_oliveira2016} to be $1.3 \pm 0.4$ Gyr, consistent with the previous chromospheric age determination by \citet{anderson_three_2014} using the activity-age relation of \citet{mamajek_hillenbrand2008} with a value of $\sim$0.8 Gyr and the gyrochronological age of 0.73--1.10 Gyr using the \citet{barnes2007} calibration. We also used a more recent tool, \texttt{gyro-interp} \citep{angus_gyro_2019}, to derive the geochronological age using the stellar rotation period and effective temperature, which led to an age of $2.6 \pm 0.2$ Gyr. However, this older age is at the upper limit of the validated calibration range and is independent of the metallicity, which is known to be a tracer of stellar ages \citep{amard_age_2020}. We, therefore, decide to converge on a stellar age ranging from 1 to 2.6\,Gyr.
\section{Photometry}\label{sec:photometry}

\subsection{Simultaneous photometry to NIRPS}
In parallel to NIRPS observations, simultaneous photometry was scheduled with the Euler (\citealt{lendl_wasp-42_2012} for more details on EulerCam) and ExTrA \citep{bonfils_extra_2015} telescopes to identify and mitigate any stellar effects like spot occultations/flares and to refine the orbital parameters. Euler is a 1.2-meter telescope at ESO's La Silla site hosting a 4k $\times$ 4k CCD camera: EulerCam. One full transit was observed on the night of 2023-07-28 with a Gunn-$r^{'}$ filter and an exposure time of 30 seconds.  The raw images were corrected for bias, flat and overscan using the standard EulerCam reduction pipeline. The ExTrA facility consists of three 0.6-meter telescopes feeding a near-infrared, multi-object spectrograph. We observed two transits on 2023-06-23 and 2023-07-28 with all three telescopes. We used the low-resolution mode (R$\sim$ 20) with an exposure time of 60 seconds. The observations were reduced using a custom reduction pipeline \citep{cointepas_2021}. No simultaneous photometry was obtained for the transit of 2023-08-24. The Euler and ExTrA light curves are shown in Figure \ref{fig:photo_models}. Due to their large systematics, we discarded the light curves of ExTrA telescope number three.

Combining these observations with archival observations, we refined the system parameters as described in Section \ref{subsec:sys_params}. Furthermore, to search for occultations of active regions, the raw light curves from individual nights are fitted with a transit and active region model \citep{PTS2018, chakraborty_sage_2023}. For this, we set wide uniform priors on active regions' position, size, and temperature (priors listed in Table \ref{tab:phot_prior}). Including active regions does not improve the Bayesian Information Criterion (BIC) by more than 4 over a pure transit model. Thus, we conclude that any active region crossings do not measurably contaminate our simultaneous NIRPS observations on 2023-06-26 and 2023-07-28. Moreover, the inspection of the HARPS radial velocities in Section\,\ref{sec:orb_arch} confirms the lack of stellar contamination in our data.


\subsection{Refined system parameters}
\label{subsec:sys_params}
To refine the system parameters, we performed an analysis combining multiple transits from Euler, ExTrA, and TESS \citep{Ricker_TESS}, using the Code for Exoplanet Analysis package (CONAN; \citealp{Lendl2020}). The code utilizes transit models from \cite{Mandel&Agol}, and we set wide uniform priors on the parameters including stellar density ($\rho_{\star}$), transit depth (R$_{\rm{p}}$/R$_{\star}$), impact parameter (b), orbital period (P$_{\rm{orb}}$), and mid-transit time (T$_{0}$). To model the limb-darkening, we set Gaussian priors on the limb-darkening (quadratic) coefficients obtained from LDCU\footnote{https://github.com/delinea/LDCU} \citep{Deline_cheops_2022}. To account for correlated noise mainly arising from instrumental systematics and weather effects, we fitted a baseline model along with the pure transit model. The baseline model is constrained by iteratively fitting different models involving airmass, sky background, (X-Y) shifts, and Full Width at half Maximum (FWHM) of stellar point spread function (PSF), minimizing the BIC. We used a Matern 3/2 Gaussian process kernel for all light curves to account for any uncorrected correlated noise. Furthermore, we adjusted the planetary mass with these updated parameters and the $K_\mathrm{\star}$ from \cite{anderson_three_2014}. The obtained stellar and planetary parameters from our analysis are listed in Table \ref{table:params}.



\section{Orbital architecture}\label{sec:orb_arch}

\subsection{Data preparation}

We excluded from our Rossiter-McLaughlin effect study the last two in-transit exposures of NIRPS Visit 3 (see Section\,\ref{sec:obs} and Fig.\,\ref{fig:condtions}) and the last two exposures in HARPS-N Visit 1 due to low S/N. The data were analyzed following the Rossiter-McLaughlin ''Revolutions'' (RMR) approach \citep{bourrier_rossiter-mclaughlin_2021}, as implemented in the \textsc{antaress} pipeline (\citealt{Bourrier2024}). For the NIRPS data, we used the NIRPS DRS \texttt{S2D} spectra\footnote{We compared these CCFs outputs to those produced by the APERO alternative NIRPS pipeline, and they are in good agreement.} as this pipeline is common to HARPS and HARPS-N data, as they are all based on the ESPRESSO pipeline. Spectra of all instruments were then passed through binary weighted cross-correlation (\citealt{baranne_elodie_1996,pepe_coralie_2002}) with a numerical mask to compute cross-correlation functions (CCFs) with a step size equal to the instrument pixel width.


The HARPS and HARPS-N spectra were cross-correlated with the standard K2 mask used in their DRS, which is closest in type to that of WASP-69 (K5). Specific masks for NIRPS spectra are still in development; the large number of telluric lines in the near-infrared spectral range makes their construction more difficult. We built a custom NIRPS mask specific to WASP-69, which exploits all out-of-transit spectra of the three processed NIRPS visits. A line-by-line analysis of the spectra \citep{artigau_line-by-line_2022} was performed to identify stellar lines with optimal radial velocity (RV) information. A template for the stellar spectrum was built by averaging the out-of-transit spectra of WASP-69, aligned in the stellar rest frame. The RVs were then derived from individual lines by cross-correlating the spectrum in each exposure with the template. Based on the derived RV distribution, we identified lines that are too affected by stellar variability and telluric contamination. This was done by calculating the correlation coefficient between the line RV, Full Width at Half Maximum (FWHM), and skewness with the Barycenter Earth radial velocity (BERV). Weights on the mask line representative of the photonic error on the line positions were defined in the same way as the ESPRESSO DRS masks (\citealt{bourrier_rossiter-mclaughlin_2021}). Following the above approach, we built NIRPS masks specific to each transit dataset. We compared the results of the RM analysis performed on each dataset with CCFs built either with the visit-specific mask or common mask. No significant difference was found between the different masks, and we thus decided to use the common mask for a more homogeneous analysis.  

\subsection{Rossiter-McLaughlin Revolutions analysis}

\begin{figure}
\includegraphics[trim=0cm 0cm 0cm 0cm,clip=true,width=\columnwidth]{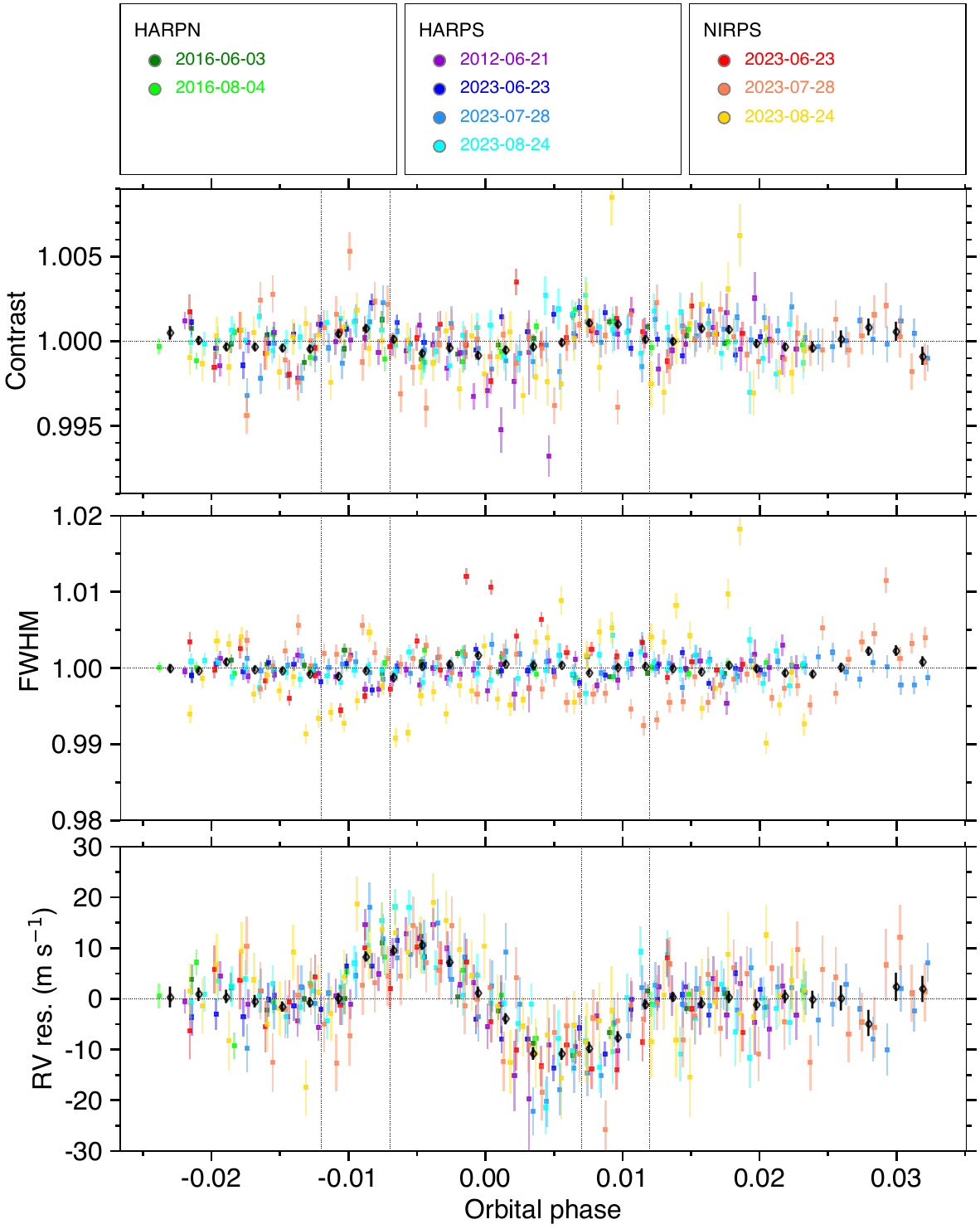}
\centering
\caption[]{Properties of the WASP-69 disk-integrated CCFs. Colored circles correspond to the best fit of the CCF in individual exposures (colored in red, blue, and green for NIRPS, HARPS, and HARPS-N, respectively). Contrast and FWHM vary over time and have been normalized to their out-of-transit mean for comparison. Plotted measurements have been binned into black diamonds to highlight the overall shape of the RM anomaly (we note that it may differ between instruments due to the different broadband limb-darkening and local stellar line shapes at optical and near-infrared wavelengths). Vertical dashed lines indicate the transit contacts.  }
\label{fig:DI_prop}
\end{figure}

We first analyzed each visit independently to assess their quality and correct for possible trends. Disk-integrated CCFs were fitted in individual exposures with a Gaussian profile to derive their contrast, FWHM, and RV. Radial velocity residuals were computed by subtracting the Keplerian model for WASP-69, informed by the properties in Table \ref{table:params}. We then searched for correlations between the out-of-transit properties and various tracers, fitting polynomials of increasing degree and adopting the model minimizing the BIC (\citealt{Schwarz1978,Kass1995,liddle_information_2007}) as the best fit. Results of this analysis are reported in Table \ref{table:detrending}. Detrending models were then computed for all exposures and applied to the CCFs as described by \cite{bourrier_polar_2022}. The data quality can be inspected in Fig.~\ref{fig:DI_prop}, which shows the properties derived from Gaussian fits to the corrected disk-integrated CCFs. The RM anomaly is clearly visible in the RV residuals, with a symmetrical shape suggesting a sky-projected spin-orbit angle close to 0$^{\circ}$. This is consistent with the W shape of the anomaly hinted at in the contrast measurements (taking into account that telluric contamination impacts the line depth dispersion), expected from a planet occulting local stellar profiles at symmetrical RV positions around mid-transit. Table\,\ref{table:compa_RVs} shows the noise properties of the out-of-transit RV residuals between instruments, showing that NIRPS is competitive with optical instruments, even for K-type stars.

\begin{table}[h]
\caption{Comparison of the out-of-transit RV residuals (Fig.~\ref{fig:DI_prop}).}
\label{table:compa_RVs}
\centering
\begin{tabular}{c c c c c}
\toprule \toprule
Instrument & Night & RMS  & Photon noise & t$\mathrm{_{exp}}$\\
 & & [m$\cdot$s$^{-1}$] & [m$\cdot$s$^{-1}$] & [s]\\
\toprule \toprule
NIRPS   & 2023-06-23 &  5.09  & 4.55  & 600  \\
NIRPS   & 2023-07-28 &  6.80  & 5.74  & 300  \\
NIRPS   & 2023-08-24 &  7.44  & 6.39  & 300  \\
HARPS   & 2012-06-21 &  3.18  & 4.33  & 300  \\
HARPS   & 2023-06-23 &  2.81  & 2.95  & 600  \\
HARPS   & 2023-07-28 &  4.28  & 4.35  & 300  \\
HARPS   & 2023-08-24 &  4.72  & 4.35  & 300  \\
HARPS-N & 2016-06-03 &  2.15  & 2.22  & 900  \\
HARPS-N & 2016-08-04 &  4.24  & 2.36  & 900  \\
\toprule
\end{tabular}
\end{table}

\begin{figure}
\includegraphics[trim=0cm 0cm 0cm 0cm,clip=true,width=\columnwidth]{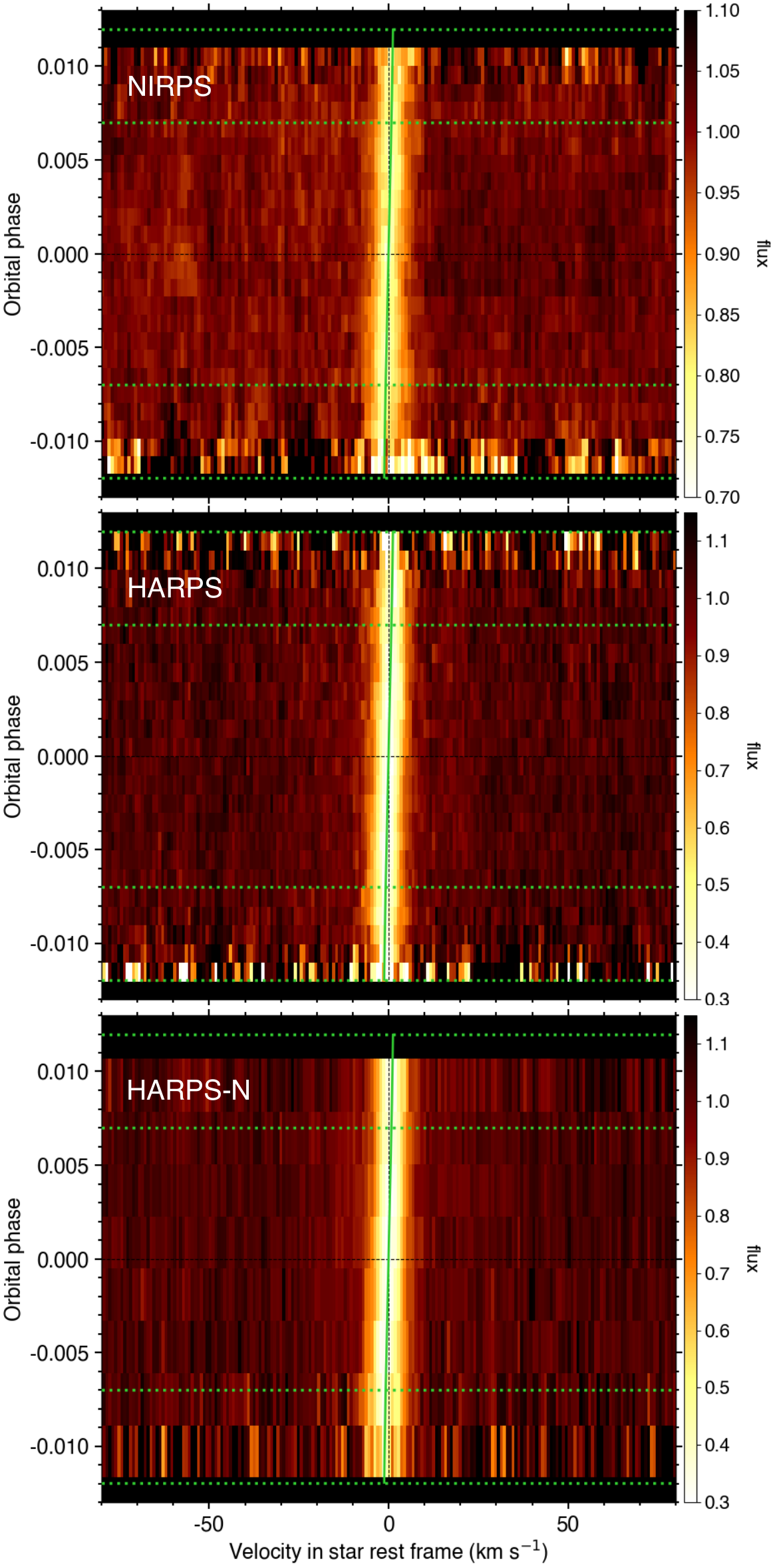}
\centering
\caption[]{Maps of WASP-69 intrinsic CCF profiles measured with NIRPS (top panel), HARPS (middle panel), and HARPS-N (bottom panel), plotted as a function of RV in the star rest frame (abscissa) and orbital phase (ordinate). Intrinsic profiles were binned together over the visits associated with each instrument for the plots. Horizontal dashed green lines show the transit contacts. The solid green line indicates the surface RVs track associated with the best RMR fit to all visits.}
\label{fig:Binned_Intr_maps}
\end{figure}

The detrended CCF time series were aligned in a common rest frame using the stellar Keplerian RV model and co-added outside of the transit to compute a master of the unocculted stellar line in each visit. Gaussian fits to these master CCFs yielded the RV shifts between the stellar and solar system barycenters, 
which were used to align all CCFs in the star rest frame. They were then scaled to their correct relative flux level using a \textit{batman} (\citealt{kreidberg_batman_2015}) light curve model, computed with the transit depth and limb-darkening properties in Table \ref{table:params}. At this stage, planet-occulted CCFs can be extracted as the difference between the master out-of-transit CCF in each visit and the CCFs in individual exposures (similarly to the reloaded RM approach, \citealt{Cegla_RM_2016}). In the RM Revolutions approach these planet-occulted CCFs are further scaled back to a common continuum to produce intrinsic CCFs (Fig.~\ref{fig:Binned_Intr_maps}), which only trace variations in the occulted stellar lines along the transit chord.  

\begin{figure}
\includegraphics[trim=0cm 0cm 0cm 0cm,clip=true,width=\columnwidth]{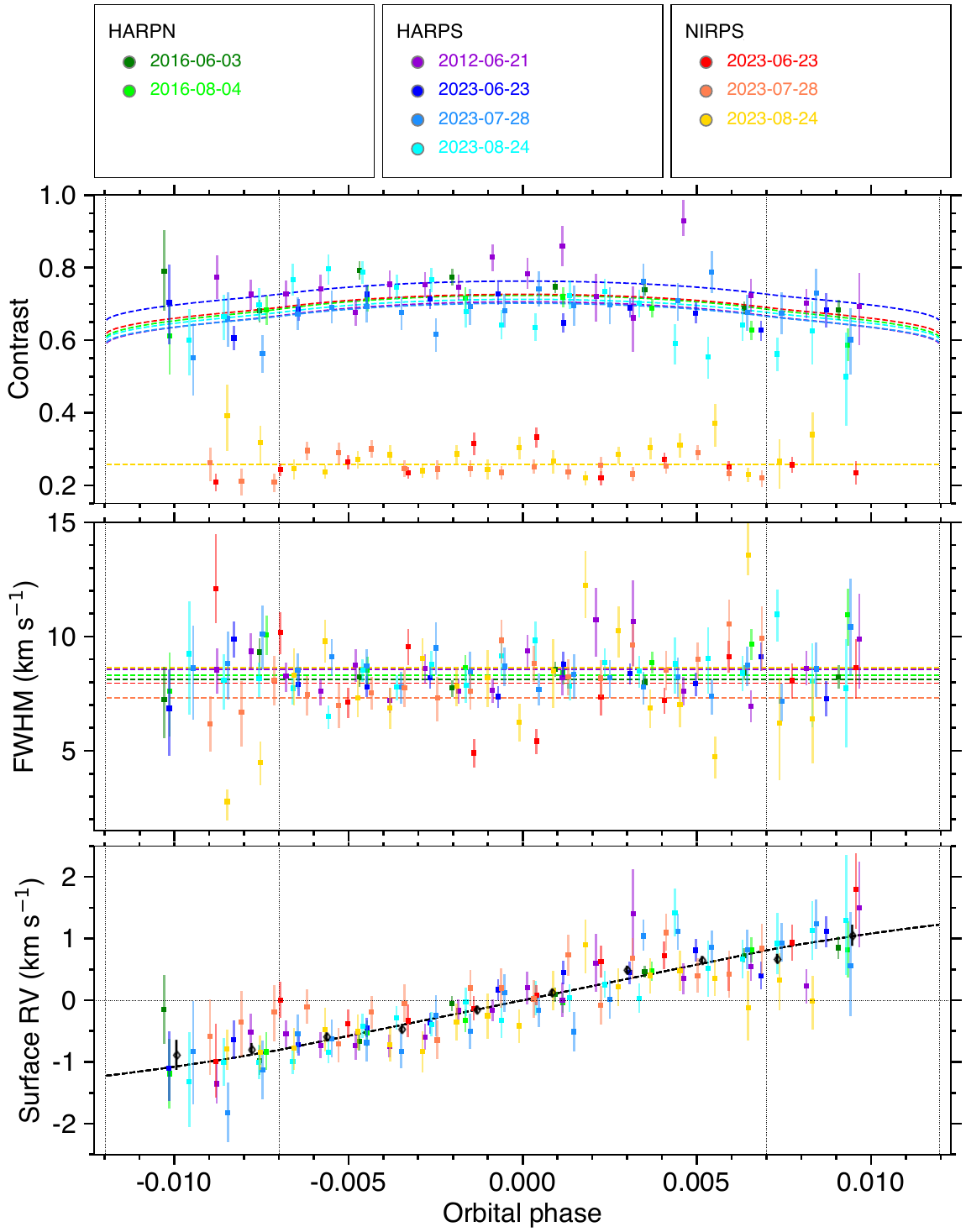}
\centering
\caption[]{Properties of the WASP-69 intrinsic CCFs along the transit chord. Colored squares correspond to the best fit to the line in individual exposures (same color scheme as in Fig.~\ref{fig:DI_prop}). Dashed-colored curves are the visit-specific contrast and FWHM models associated with the best RMR fit. The dashed black curve is the common surface RV model.  Plotted RV measurements have been binned into black circles to highlight the good agreement with the model. Vertical dashed lines indicate the transit contacts.  }
\label{fig:Intr_prop}
\end{figure}

As can be seen in Fig.~\ref{fig:Intr_maps}, the track of stellar lines from the planet-occulted regions is clearly detected in all visits. In a first step, we fitted intrinsic CCFs in individual exposures to assess their quality, identify which exposures can be used in the global RMR fit, and which models best describe the variations of the local stellar line along the transit chord. Intrinsic CCFs were fitted with Gaussian profiles convolved to a width equivalent to the resolving power of each spectrograph so that the derived contrast, FWHM, and RV trace the intrinsic stellar properties and are more comparable between instruments. We used \texttt{emcee} Monte Carlo Markov Chain (MCMC, \citealt{Foreman2013}) to sample the posterior probability distributions of the fitted parameters, taking their median as the best estimate and setting the associated uncertainty ranges to the 1-$\sigma$ Highest Density Intervals\footnote{The HDI are equivalent to the range of values that encompass 68.3\% of the PDF on each side of the median when the distribution is Gaussian.}. Broad uniform priors were set on the parameters: $\mathcal{U}$(0,1) for the contrast, $\mathcal{U}$(0, 15)\,km$\cdot$s$^{-1}$ for the FWHM, and $\mathcal{U}$(-10, 10)\,km$\cdot$s$^{-1}$ for the RV. Despite some larger dispersion in the contrast and FWHM of the NIRPS data due to residual telluric contamination, the time series do not show strong deviations from the expected trends and between visits of a same instrument. In particular, the surface RVs, which are the most constraining parameters in the RMR fit, are remarkably consistent across all datasets. The final time series used in the RMR fit (Fig.~\ref{fig:Intr_prop}) only excludes exposures at the limbs of the star, where the lower flux and surface of the occulted stellar regions yield intrinsic CCFs that are too noisy to be constraining. 

Polynomials of the sky-projected distance to the star center were fitted to the intrinsic contrast and FWHM time series to determine the optimal degree of the model and possible variations of its coefficients between epochs, again using the BIC for model comparison. The NIRPS local stellar lines are best modeled with a constant contrast common to all epochs and a constant FWHM specific to each epoch. The HARPS lines are best modeled with a linear contrast variation common to all epochs, modulated by a contrast level specific to each epoch, and a constant FWHM common to all epochs. The HARPS-N lines are best modeled with a linear contrast variation common to all epochs, modulated by a contrast level and a constant FWHM, both specific to each epoch. The HARPS/HARPS-N contrast and FWHM series are fairly similar between epochs, even though the first HARPS dataset was taken more than 10 years earlier. Considering that the same CCF mask was used for all-optical data and that the HARPS and HARPS-N spectra cover roughly the same spectral range, this suggests that the photosphere of WASP-69 is quite stable over time. The surface RV time series are consistent between the three instruments' visits and are best modeled with solid-body rotation, with no differential rotation or convective blueshift allowed by the data. We note that the antisymmetry of the surface RVs around mid-transit already indicates a sky-projected angle close to 0$^{\circ}$.

In a second step, we fitted the full intrinsic CCF profiles over the joint series of selected exposures. This RMR approach allows for the exploitation of the full information contained in the transit time series. Intrinsic Gaussian stellar lines are computed for each exposure using contrast, FWHM, and RVs defined by the models identified in the first step. The line properties are brightness-averaged over a numerical grid, tiling the regions occulted by the planet during each exposure, naturally accounting for the blur induced by the planet's motion. Theoretical line models are convolved with the instrumental response of each instrument before they are compared with the data. The fit is constrained enough that uniform, uninformative priors were set on all parameters, whose Posterior Distribution Functions (PDFs) were again sampled using \texttt{emcee}.

\begin{table}[h]
\caption{Comparison of RM results between instrument datasets.}
\label{table:compa_RM}
\centering
\begin{tabular}{c c c}
\toprule \toprule
Instrument & $v sin i_\mathrm{\star}$ [km$\cdot$s$^{-1}$] & $\lambda$ [deg] \\
\toprule \toprule
NIRPS &  1.34$\pm$0.12   & 2.1$^{+2.7}_{-2.5}$  \\
HARPS  &   1.60$\pm$0.08  &   0.2$\pm$1.4   \\
HARPS-N &  1.59$\pm$0.12   &  -1.4$\pm$2.1   \\ 
All &   1.54$\pm$0.05  &   0.05$\pm$1.1  \\ 
\toprule
\end{tabular}
\end{table}

We first performed RMR fits on the joint datasets from each instrument (Table~\ref{table:compa_RM}). Projected spin-orbit angles are consistent between all datasets. Projected velocities derived from the optical HARPS and HARPS-N datasets are the same and only marginally larger than the value derived from the near-IR NIRPS datasets. Our results for HARPS-N are consistent with the values derived by \citet{casasayas-barris_detection_2017} from a classical analysis of the RM anomaly in the same datasets, with $\lambda$ = 0.4$^{+2.0}_{-1.9}$\,$^{\circ}$ and $v sin i_\mathrm{\star}$ = 1.57$^{+0.13}_{-0.07}$\,km$\cdot$s$^{-1}$ (derived from their angular rotation parameter).

\begin{figure}
\includegraphics[trim=0cm 0cm 0cm 0.5cm,clip=true,width=\columnwidth]{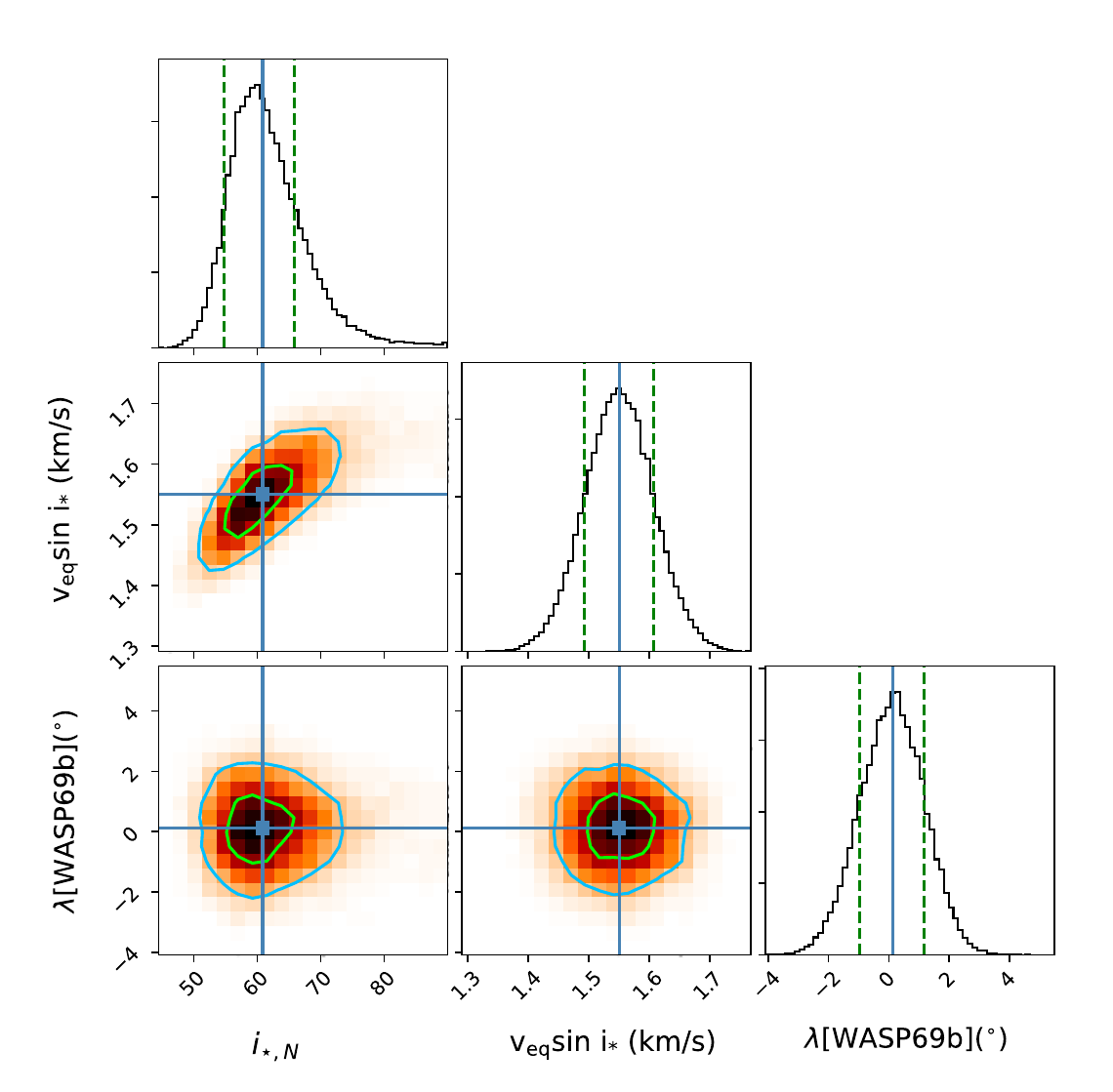}
\centering
\caption[]{Correlation diagrams for the PDFs of the stellar inclination (Northern configuration), sky-projected stellar rotational velocity, and sky-projected spin-orbit angle, as fitted or derived from our final RMR fit (see text). Green and blue lines show the 1 and 2$\sigma$ simultaneous 2D confidence regions that contain, respectively, 39.3\% and 86.5\% of the accepted steps. 1D histograms correspond to the distributions projected on the space of each line parameter, with the green dashed lines limiting the 68.3\% HDIs. The blue lines and squares show the median values.}
\label{fig:Corr_diag}
\end{figure}

We then performed a joint RMR fit to all datasets simultaneously and showed the residuals from the best-fit models in Fig.~\ref{fig:Intr_maps}. Residual features, likely due to telluric lines, are still visible in the NIRPS datasets. Nonetheless, they are mainly located outside of the transit or far from the planet-occulted lines and have little impact on our analysis. This is supported by the good agreement between the values derived from the NIRPS and other datasets. The final results for the system are reported in Table \ref{table:params}, with correlation diagrams for some of the model parameters shown in Fig.~\ref{fig:Corr_diag}. The corresponding theoretical time series for the intrinsic line properties are overplotted with measured values in Fig.~\ref{fig:Intr_prop}.

\begin{figure}
\includegraphics[trim=0cm 0cm 0cm 0.0cm,clip=true,width=\columnwidth]{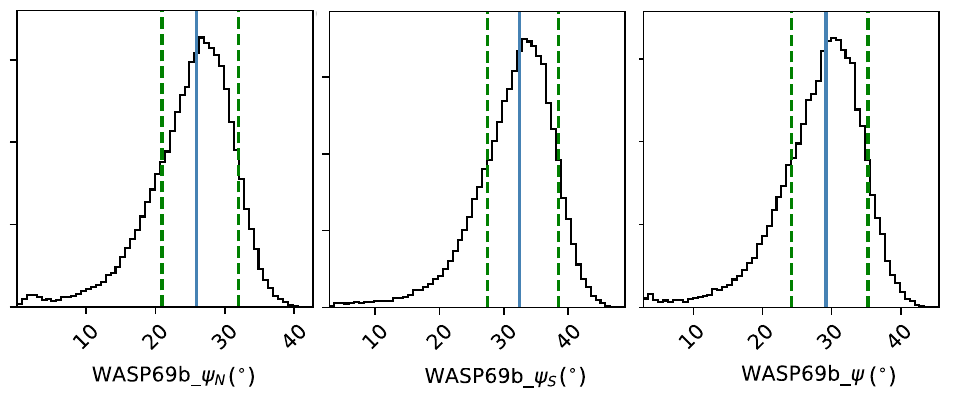}
\centering
\caption[]{PDFs of the 3D spin-orbit angle in the Northern, Southern, and combined configurations. Green dashed lines limit the 68.3\% HDIs. Blue lines show the median values.}
\label{fig:Psi_PDFs}
\end{figure}

\begin{figure}
\includegraphics[trim=0cm 3cm 0cm 1cm,clip=true,width=\columnwidth]{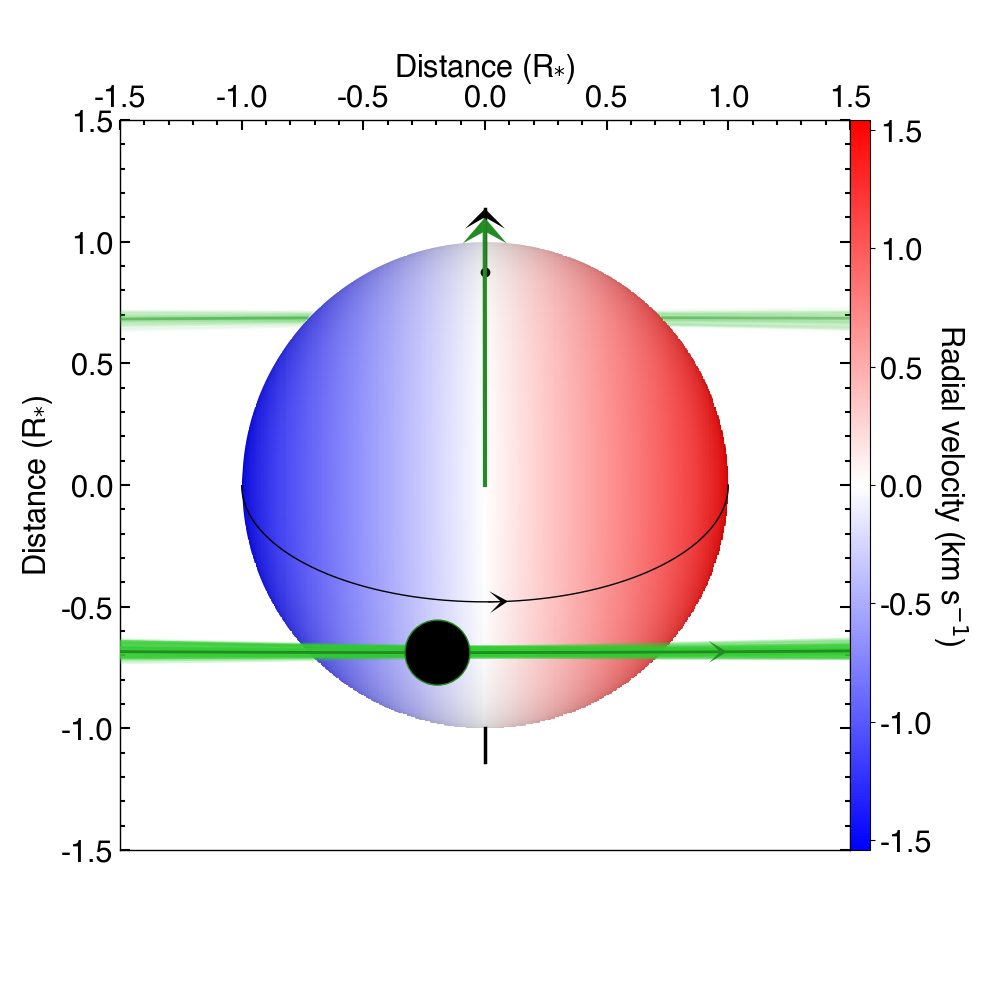}
\centering
\caption[]{Projection of the WASP-69 system on the sky plane for the best-fit orbital architecture. We show the configuration where the stellar spin axis (shown as a black arrow extending from the north pole) is pointing toward the Earth. The stellar equator is plotted as a solid black line. The stellar disk is colored as a function of its surface RV field. The normal to the planetary orbital plane is shown as a green arrow extending from the star center. The green solid curve represents the best-fit orbital trajectory, surrounded by thinner lines showing orbits obtained for orbital inclination, semi-major axis, and sky-projected spin-orbit angle values drawn randomly within 1$\sigma$ from their probability distributions. The star, planet (black disk), and orbit are to scale.}
\label{fig:RM_3D_view}
\end{figure}

This is the first RM analysis exploiting as many as nine datasets together, from six individual transits, and we reach exquisite precisions of 50\,m$\cdot$s$^{-1}$ and 1$^{\circ}$ on the sky-projected rotational velocity and spin-orbit angle, respectively. We used this exquisite temporal baseline to search for stellar obliquity precession by looking at the time series of the spin-orbit angle. However, we do not observe any obvious trend and the RMS of the obliquity is in agreement with the 1-$\sigma$ uncertainty. Despite a decade of observations, the baseline is too short to observe evidence of stellar precession. Further constraints on the system can be derived by using independent constraints, especially knowledge of the stellar equatorial rotation period from photometry. We ran again our final MCMC fit using the independent variables $R_\star$, $P_\mathrm{eq}$ and $\cos i_\star$ as jump parameters instead of $v_\mathrm{eq} \sin i_\star$ (see \citealt{Masuda2020} and \citealt{bourrier_dream_2023}). We set a uniform prior on $\cos i_\star$, and Gaussian priors from measured values on $R_\star$ and $P_\mathrm{eq}$  (Table \ref{table:params}, from \citealt{anderson_three_2014}). We then derived from the results the PDFs on the stellar inclination $i_\star$ and 3D spin-orbit angle $\psi$: 
\begin{equation}
\label{eq:cospsi}
    \psi = \arccos\left( \sin i_\star \sin i_\mathrm{p} \cos \lambda + \cos i_\star \cos i_\mathrm{p}\right).
\end{equation}
The fit only remains sensitive to $\sin i_\star$, so that there are two degenerate ``northern'' ($\psi_{\rm N}$ = 25.9$^{+6.1}_{-5.0}$ $^{\circ}$) and ``southern'' ($\psi_{\rm S}$ = 32.5$^{+6.1}_{-5.0}$ $^{\circ}$) configurations corresponding respectively to $i_\star$ and $\pi - i_\star$. Since the PDFs for $\psi_{\rm N}$ and $\psi_{\rm S}$ are similar and can be considered equiprobable, we also combined them to yield $\psi$ = 29.2$^{+6.1}_{-5.0}$ $^{\circ}$ (Fig.~\ref{fig:Psi_PDFs}). These results reveal that the WASP-69 system is, in fact, moderately misaligned (Fig.~\ref{fig:RM_3D_view}). This illustrates the observational bias highlighted by \citet{attia_dream_2023} that systems considered aligned based on their sky-projected spin-orbit angle are more likely to be misaligned. This result is strengthened by the consistency between our measurement for $i_\star$ = 60.8$^{+5.0}_{-6.1}$ $^{\circ}$ and the value of about 68$^{\circ}$ favored by the independent analysis of spot modulation in TESS data by \citet{chakraborty_sage_2023}. 

\section{Helium triplet}\label{sec:helium}
Hereafter, in the analysis of the helium triplet, we use the three NIRPS transits reduced with the APERO pipeline, as it has slightly fewer systematics compared to the NIRPS-DRS (see Appendix\,\ref{app:pipelines_comp}).

\subsection{Data analysis}\label{subsec:data_analysis}

We used the same data analysis procedures presented by \cite{allart_homogeneous_2023}, which is based on a standard algorithm developed to study exoplanet atmospheres at high resolution \citep[e.g.,][]{wyttenbach_spectrally_2015,casasayas-barris_detection_2017,allart_spectrally_2018,allart_high-resolution_2019,seidel_hot_2020} and is described below.\\

We focus the analysis on \'echelle order 134 (order 15 for both pipelines covering 10737.86-10888.24\,\AA) from the \texttt{e2ds} spectra, where the helium triplet falls\footnote{\'Echelle order 133 also includes the spectral domain around the helium triplet on its left edge but at very low S/N due to the drop of flux from the blaze. We decided to exclude this order from the analysis.}. The telluric-corrected stellar spectra are shifted to the stellar rest frame using the systemic velocity measured in Section\,\ref{sec:orb_arch}, normalized using the median flux in two bands (10823-10829\,\AA\, and 10836-10842\,\AA), and remaining outliers (e.g., cosmic rays) are sigma-replaced following \cite{allart_search_2017}. To build the stellar reference spectrum, hereafter called master-out spectrum, we combine all spectra obtained before phase -0.0149 and spectra obtained after phase 0.250 --- these spectra are hereafter called out-of-transit spectra. Spectra obtained between these phases and the transit's ingress and egress are not included in the out-of-transit spectra as we observe clear excess absorption that could contaminate the master-out. We caution that defaulting to using out-of-transit spectra based on the optical transit light curve when an extended helium signature is present would introduce biases in the analysis. Figure\,\ref{fig:master_tell} displays the master-out of WASP-69 for each visit before and after applying the telluric correction.\\

\begin{figure}
\includegraphics[width=\columnwidth]{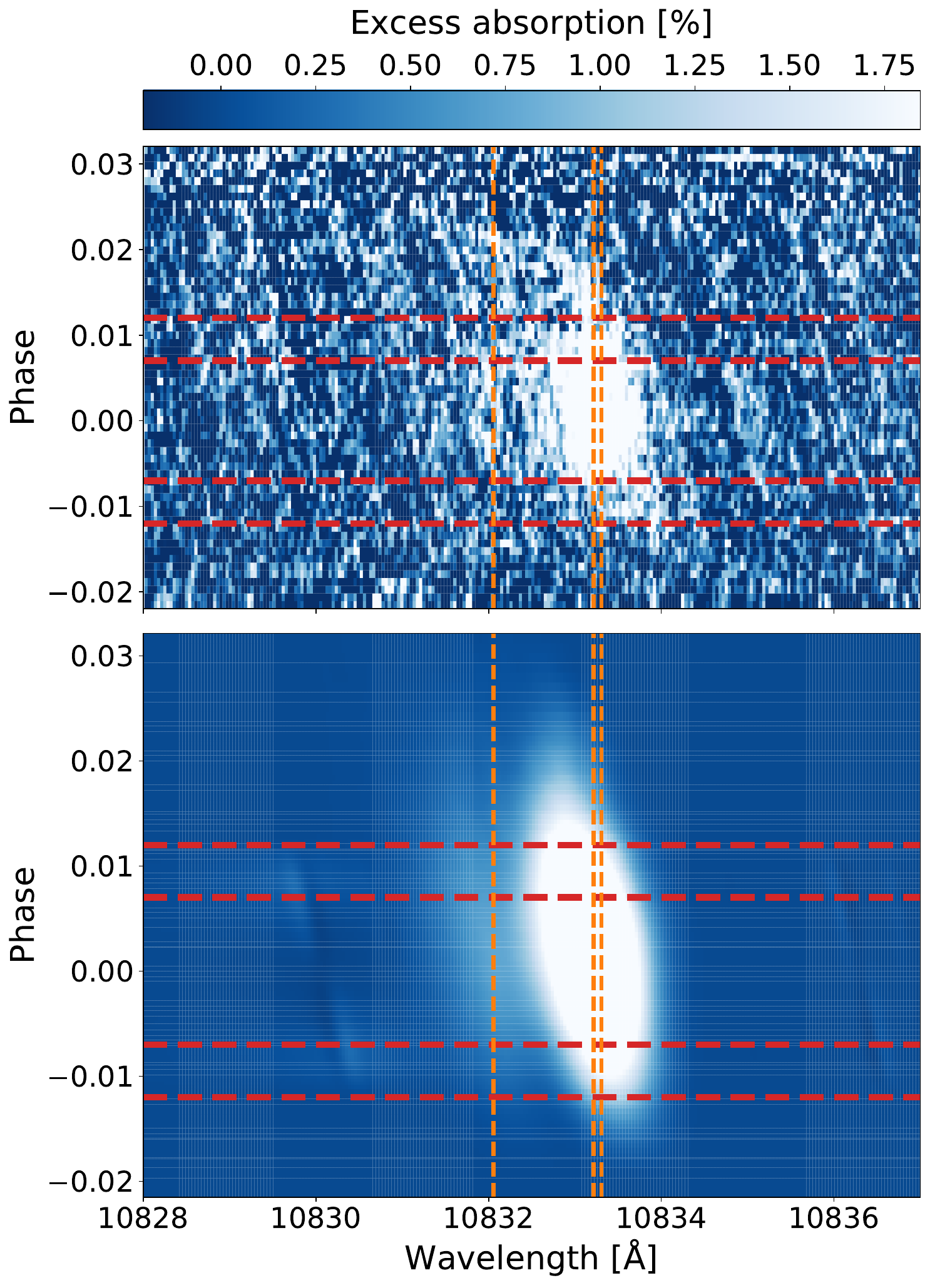}
\centering
\caption[]{Transmission spectroscopy map of WASP-69b in the planet rest frame. The \textit{top} panel displays the data, while the \textit{bottom} panel is the best-fit model from the thermosphere and exosphere. The truncated color scale shows excess absorption in white. The red dashed horizontal lines are the contact lines from bottom to top: t$_1$, t$_2$, t$_3$, and t$_4$. The vertical orange dashed lines indicate the helium lines' positions.}
\label{fig:TS_map}
\end{figure}

The next step is to remove stellar features to obtain a transmission spectroscopy map. We thus divide each spectrum of the time series by the master-out and then Doppler-shift them to the planet rest frame based on the parameters in Table\,\ref{table:params}. Figure\,\ref{fig:TS_map} displays the averaged transmission spectroscopy map over the three transits of WASP-69\,b in the planet rest frame. We observe a clear excess absorption signature during transit, mostly following the expected position of the planetary track.\\

\begin{figure}
\includegraphics[width=\columnwidth]{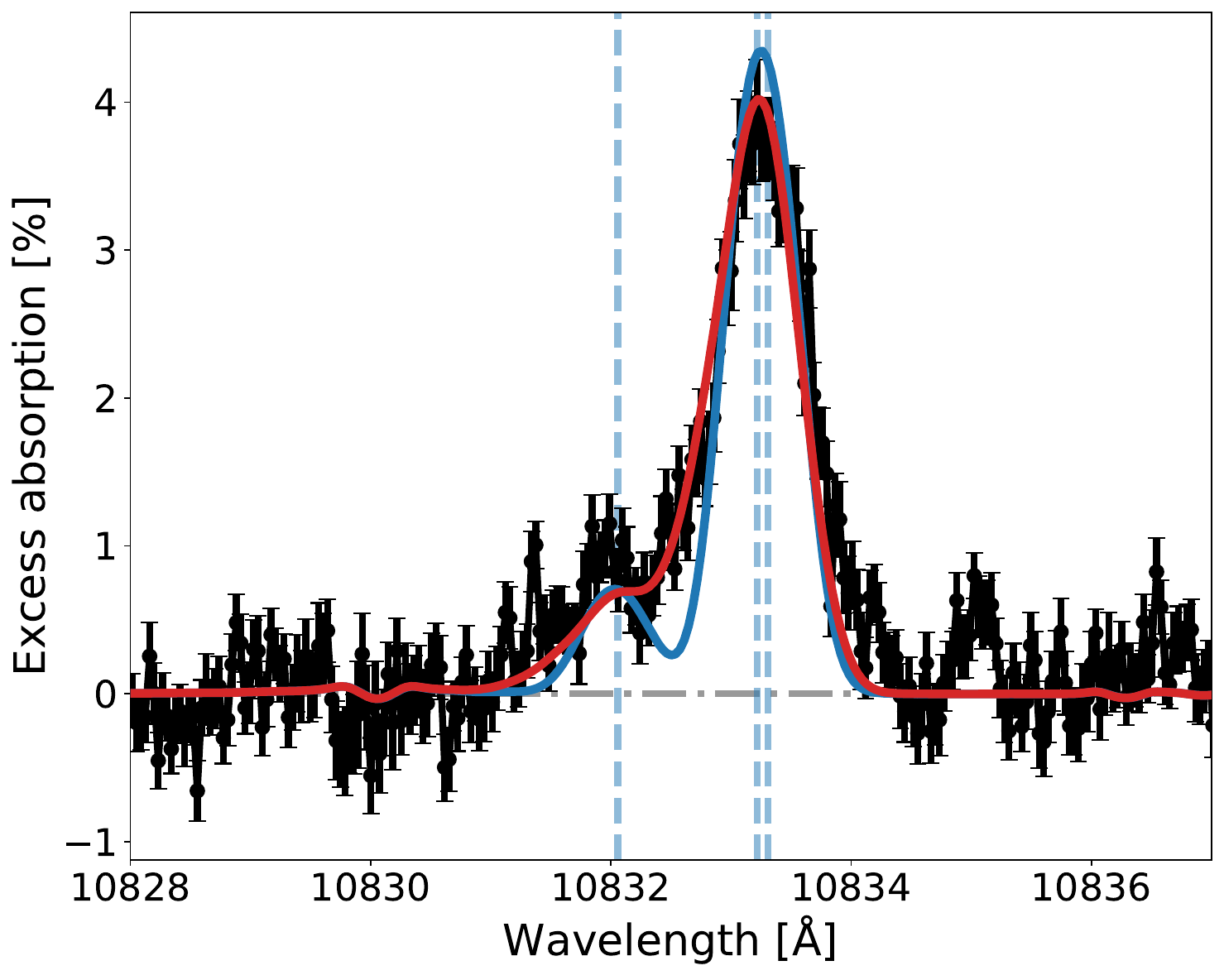}
\centering
\caption[]{Average transmission spectrum of WASP-69\,b in the planet rest frame in black. The blue curve is the best-fit model from the thermosphere only, while the red curve is the best-fit model from the thermosphere and exosphere. Spectra between t$_1$ and t$_4$ were used to built the transmission spectrum. The vertical blue dashed lines indicate the helium lines' positions. The bump at 10 835\,\AA\ is likely due to telluric residuals.}
\label{fig:TS}
\end{figure}

The 1D transmission spectrum is computed for each night as the average of the transmission spectroscopy map weighted by a modeled white light curve \citep{allart_high-resolution_2019,allart_homogeneous_2023,mounzer_hot_2022}. The \texttt{batman} \citep{kreidberg_batman_2015} package was used to model the white light curve with the parameters from Table\,\ref{table:params}, where the quadratic limb darkening coefficients have been estimated in the $J$-band with the \texttt{EXOFAST} \citep{eastman_exofast_2013} code based on the tables of \cite{claret_gravity_2011}. This scaling is necessary to properly consider the true contribution of the ingress and egress spectra into the transit average transmission spectrum. Finally, the transmission spectra are averaged, with weights set by their uncertainties, to build the average transmission spectrum. Figure\,\ref{fig:TS} displays the average transmission spectrum of WASP-69\,b around the helium triplet measured during transit. We observe a clear excess absorption of 3.17$\pm$0.05\% over the 0.75\,\AA\ passband centered at 10833.22\,\AA\ with a maximum absorption of 4.02\%. This signature is quite symmetric but slightly blueshifted with respect to the main helium line, while even the weakest component of the helium triplet is visible at 10832.06\,\AA.

\begin{figure}
\includegraphics[width=\columnwidth]{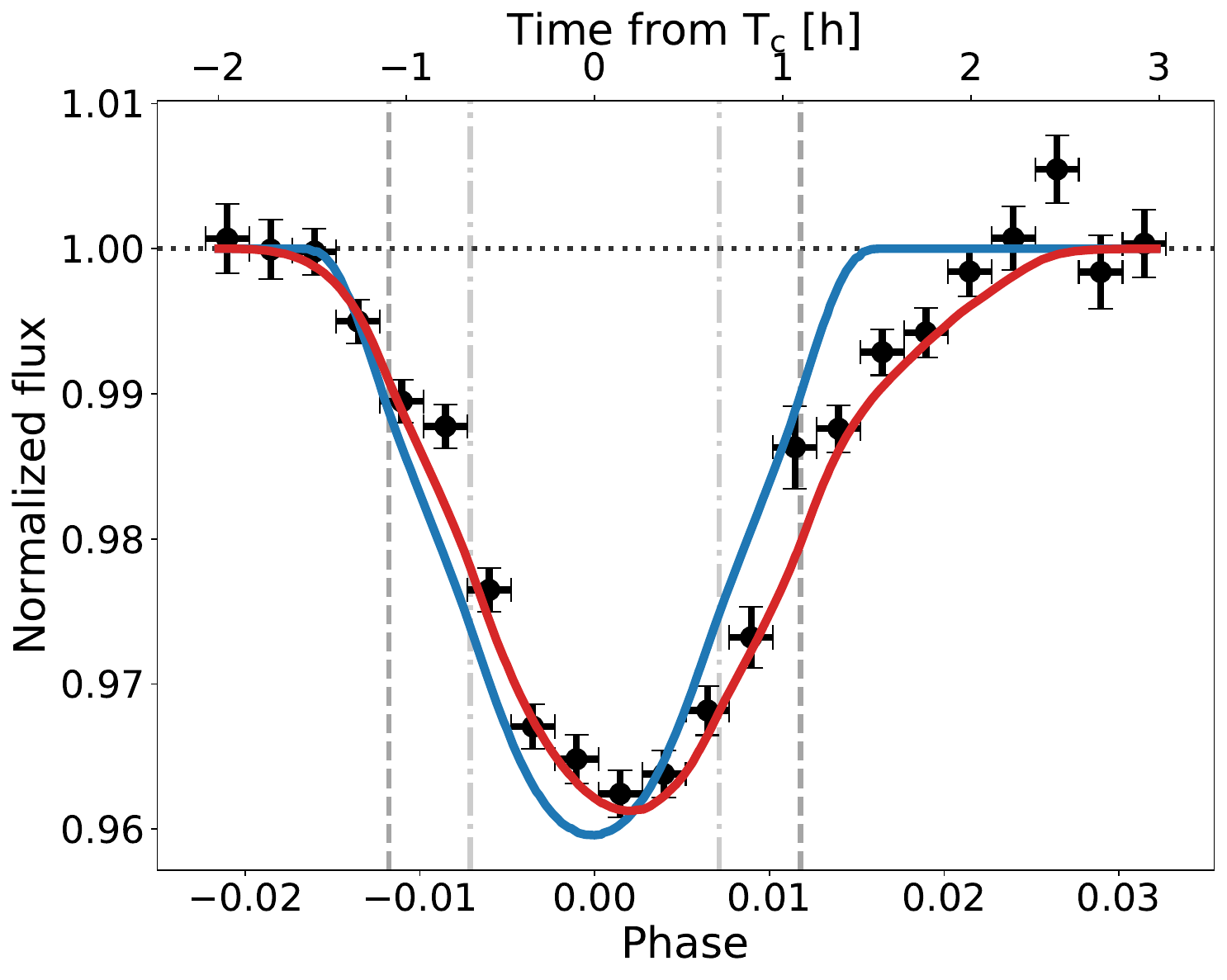}
\centering
\caption[]{Average excess helium light curves of WASP-69\,b in black. The blue curve is the best-fit model from the thermosphere only, while the red curve is the best-fit model from the thermosphere and exosphere. The gray dashed vertical lines are the contact lines from left to right: t$_1$, t$_2$, t$_3$, and t$_4$. The grey horizontal dotted line is the continuum level.}
\label{fig:LC}
\end{figure}

We compute the helium light curve to study the temporal variability of the signal by measuring the excess absorption, assuming a symmetric signal, in a passband of 0.75\,\AA\ centered at 10833.22\,\AA\ for each exposure of the transmission spectroscopy map in the planet rest frame. Figure\,\ref{fig:LC} displays the measured averaged helium light curve for WASP-69b, where we observe an asymmetric light curve with the maximum absorption just after mid-transit and clear pre- and post-transit absorption. Moreover, the measured excess absorption is larger during egress than ingress, and clear post-transit absorption is detected up to phase 0.021. This indicates trailing material behind the planet up to 50 minutes after the end of the transit. We note that no excess absorption is detected between phases 0.021 and 0.025, even if the spectra are not used in the reference master-out. Table\,\ref{tab:LC_abs} summarizes the excess absorption pre-transit, during ingress, mid-transit, egress, and post-transit (defined as t$_4$ to phase 0.021).

\begin{table}
    \centering
    \caption{Excess absorption measured on the helium light curve for different transit phases: ingress, mid-transit, egress, and post-transit.}
    \begin{tabular}{r c}
        \toprule \toprule
        Phase & Excess absorption [\%] \\
        \toprule \toprule
        -0.021 -- t$_1$& 0.17$\pm$0.09 \\
        t$_1$ -- t$_2$ & 1.13$\pm$0.11 \\
        t$_2$ -- t$_3$ & 3.26$\pm$0.07 \\
        t$_3$ -- t$_4$ & 2.22$\pm$0.17 \\
        t$_4$ -- 0.021 & 0.86$\pm$0.09 \\
        0.021 -- 0.033 & -0.04$\pm$0.10 \\
        \toprule
    \end{tabular}
    \label{tab:LC_abs}
\end{table}

\begin{figure}
\includegraphics[width=\columnwidth]{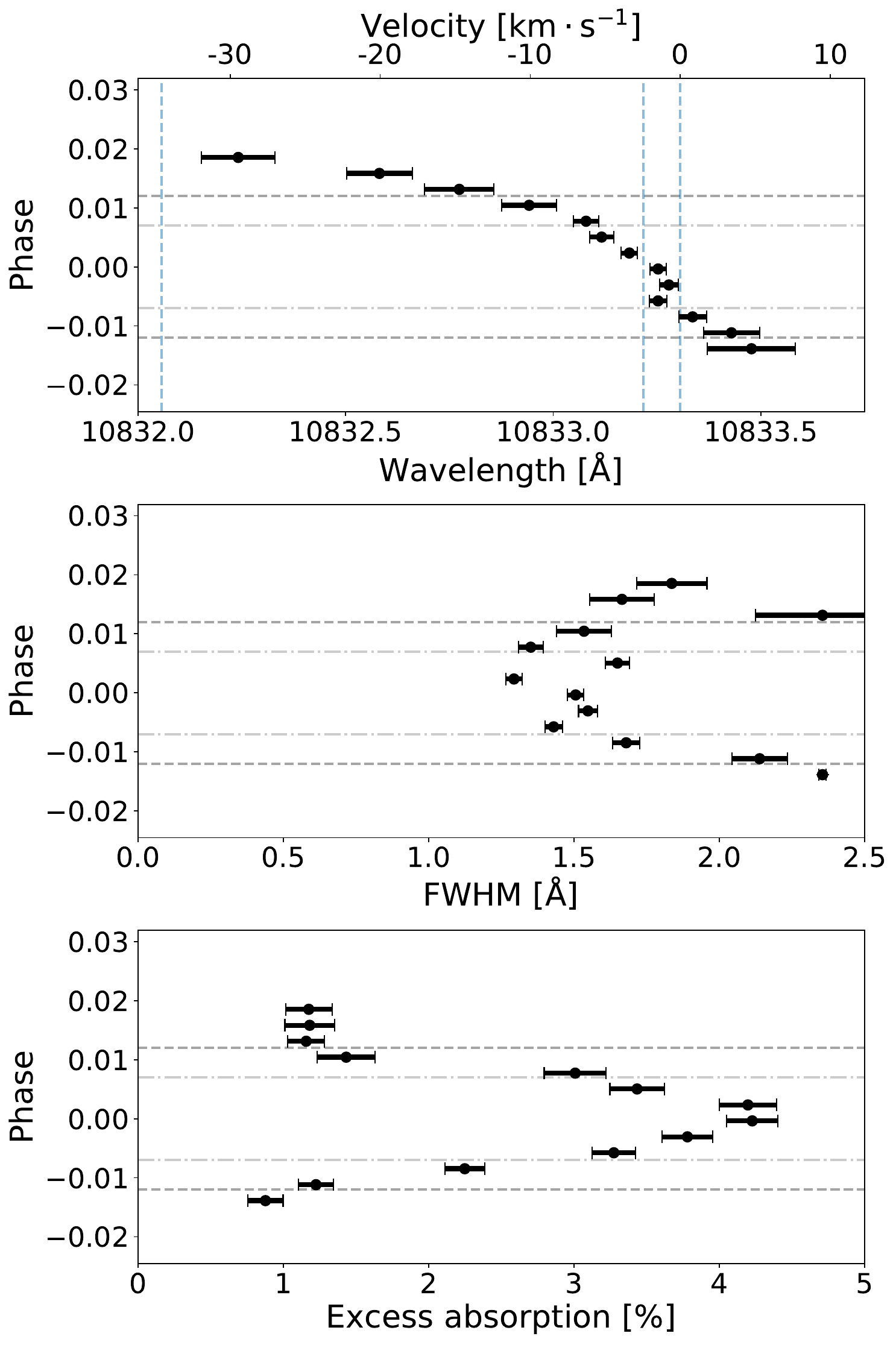}
\centering
\caption[]{Properties of the Gaussian fit on the helium signature time series binned by 15 minutes. \textit{Top:}  Gaussian position as a function of phase. The equivalent in velocity relative to the main lines of the triplet is indicated as the top label. \textit{Middle:} Gaussian Full Width at Half Maximum as a function of phase. \textit{Bottom:} Gaussian amplitude (equivalent to excess absorption) as a function of phase. The gray dashed horizontal lines are the contact lines from bottom to top: t$_1$, t$_2$, t$_3$, and t$_4$.}
\label{fig:TS_param}
\end{figure}

We then binned the transmission spectroscopy time series in bins of 15 minutes. We applied a Gaussian fit to the individual transmission spectra where excess absorption is detected between phases 0.0149 and 0.250 (based on Fig.\,\ref{fig:LC}) to estimate the helium signature's centroid, FWHM, and amplitude. Figure\,\ref{fig:TS_param} displays those parameters' temporal evolution. First, we notice that the amplitudes of the Gaussians confirm the helium triplet light curve. Second, we measure an average FWHM of $\sim$1.6\,\AA\, equivalent to $\sim$43\,km$\cdot$s$^{-1}$, with a minimum during mid-transit of 1.3\,\AA\ and a maximum during ingress and egress of 2.3\,\AA. Finally, we detect a clear velocity shift of the helium signature from 4.74$\pm$2.93\,km$\cdot$s$^{-1}$ before ingress to $-$1.48$\pm$0.55\,km$\cdot$s$^{-1}$ at mid-transit, $-$14.73$\pm$2.30\,km$\cdot$s$^{-1}$ at egress, and up to $-$29.46$\pm$2.46\,km$\cdot$s$^{-1}$ 50 minutes after egress. This variation is steady during the first half of the transit, increases slowly in the second half, and extends rapidly to high velocities after transit. This is indicative of dynamical outflows in the upper atmosphere of WASP-69\,b.

\subsection{Analysis of individual nights }\label{subsec:night_analysis}
\begin{figure}
\includegraphics[width=\columnwidth]{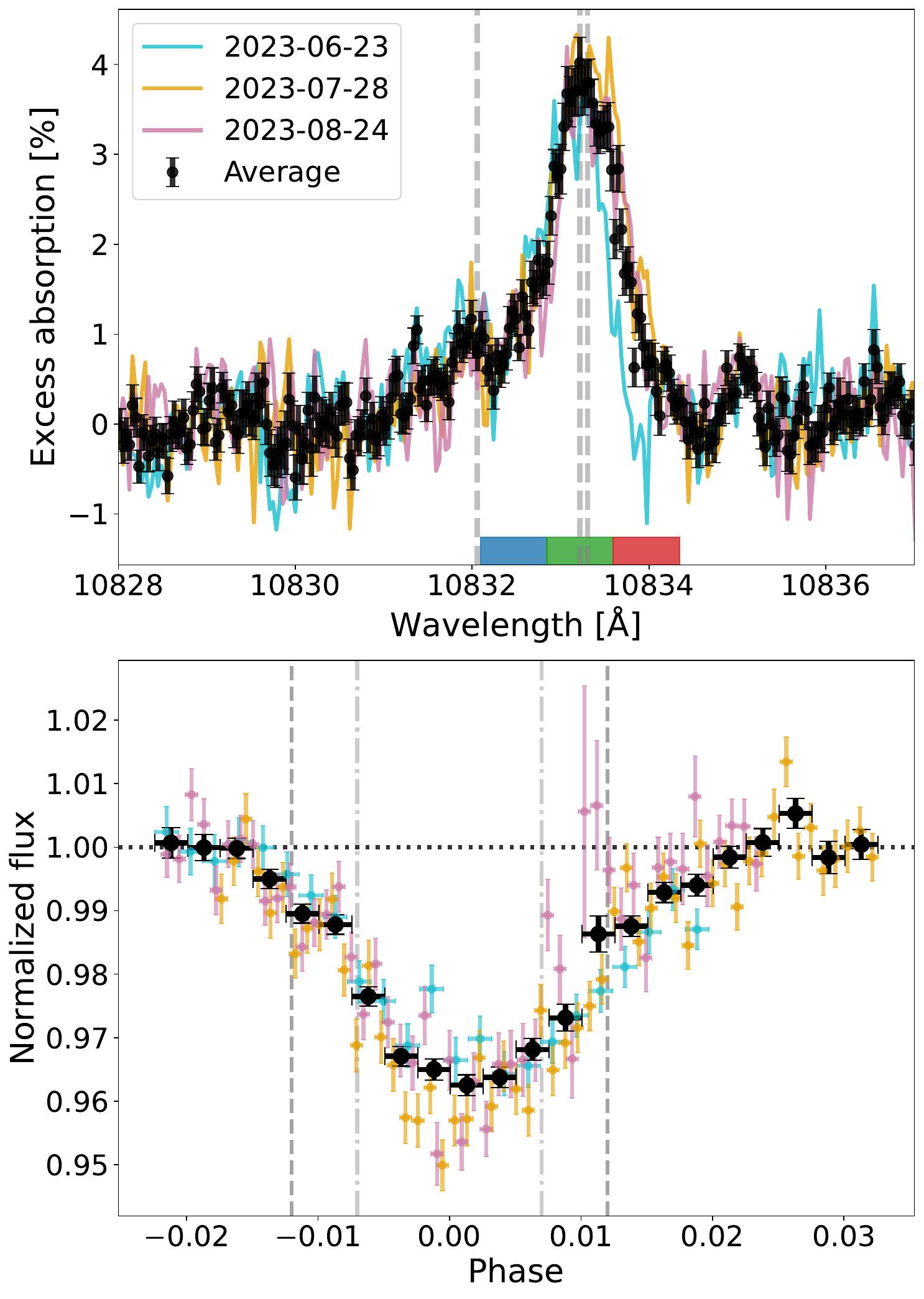}
\centering
\caption[]{Transmission spectra (\textit{top}) and excess helium light curves (\textit{bottom}) comparison between transits (in cyan, orange, and pink) and the average in black. On the top panel, three spectral regions (blue, green, and red) are identified to measure the temporal variation of the helium signature.}
\label{fig:TS_LC_night}
\end{figure}

Figure\,\ref{fig:TS_LC_night} compares each transit transmission spectrum and helium light curve with their average. From the light curve, the three transits are in good agreement except for the loss of flux during the egress of the 2023-08-24 visit (see Section\,\ref{sec:obs}). The transmission spectra are also in overall good agreement between each night with similar maximum excess absorption. However, we note some difference in the shape of the helium triplet, which is confirmed when computing the excess absorption in three 0.75\,\AA\ passbands centered at 10832.47, 10833.22 and 10833.97\,\AA\ (Table\,\ref{tab:Excess_abs}). The three transits in the bluest spectral domain agree within 2.1$\sigma$ of the average transmission spectrum. In the central domain (green band in the top panel of Fig.\,\ref{fig:TS_LC_night}), they disagree up to 3.8$\sigma$ with the average transmission spectrum and up to 6.3$\sigma$ between the 2023-06-23 and 2023-07-28 transits. In the reddest spectral domain, there is a discrepancy of up to 6.7$\sigma$ between the individual nights and their average, and about 10.7$\sigma$ between the 2023-06-23 and 2023-07-28 transits. Telluric residuals cannot explain the variations observed in the helium profile over the three transits as they are at redder wavelengths (Fig.\,\ref{fig:master_tell}). \cite{allart_homogeneous_2023} proposed that inhomogeneities on the stellar surface mimicking observable signals could induce a fraction of the signal they observed. We thus used the same model as they described based on \cite{andretta_estimates_2017}. This toy model assumes that the stellar surface is split into two parts, and one of them contains all the stellar helium absorption signal. Using a filling factor of $\sim$0.7 with all the stellar helium absorption coming from the dark region ($\alpha$=0), we estimate that the maximum signal coming from the star is about 0.57\% and 0.13\% in the central and reddest spectral domains, respectively. While pseudo-stellar variability can explain the discrepancy observed in the central passband, it cannot reproduce the reddest passband discrepancy. Moreover, the EWs of the helium stellar line from 10832.8 to 10833.8\,\AA\ measured on the master-out of each transit are within the uncertainties with a slightly more discrepant value for the 2023-06-23 visit. This is in agreement with the spot coverage we can expect for the last two visits (2023-07-28 and 2023-08-24), which are within 1.2 stellar rotations, while the first visit (2023-06-23) is at more than 1.5 stellar rotations of the second visit. Therefore, the observed variability between transits cannot be explained with a variable stellar pseudo-signature alone, but a fraction of this variability likely comes from stellar contamination. Consequently, planetary variability remains the main hypothesis to explain the observed variability. Upcoming observations from the NIRPS WP3 program will help us investigate the temporal variability of WASP-69\,b atmosphere in more detail.

\begin{table}
    \centering
    \caption{Excess absorption measured on the transmission spectrum using a 0.75\,\AA\ passband around three central wavelengths for each transit and for the average transmission spectrum.}    \begin{tabular}{lccc}
        \toprule \toprule
        \multirow{2}{*}{Night} & \multicolumn{3}{c}{Central wavelength [\AA]} \\
            & 10832.47 & 10833.22 & 10833.97 \\
        \toprule \toprule
        2023-06-23 [\%] & 1.28$\pm$0.08 & 2.84$\pm$0.10 & 0.36$\pm$0.09 \\
        2023-07-28 [\%] & 1.03$\pm$0.07 & 3.47$\pm$0.08 & 1.32$\pm$0.07 \\
        2023-08-24 [\%] & 1.04$\pm$0.09 & 3.03$\pm$0.10 & 1.10$\pm$0.11 \\
        Average    [\%] & 1.11$\pm$0.04 & 3.17$\pm$0.05 & 0.96$\pm$0.05 \\
        \toprule
    \end{tabular}
    \label{tab:Excess_abs}
\end{table}

\subsection{Interpreting the helium signal}\label{subsec:model_he}

We generated synthetic observations of the WASP-69\,b transits using the EVaporating Exoplanets (EVE) code \citep{bourrier_3d_2013,bourrier_evaporating_2016}. EVE simulates the planetary atmosphere and the system architecture in 3D, naturally accounting for geometrical effects on the absorption spectra. The code generates a time series of disk-integrated spectra at high temporal and spectral resolution by simulating the planet's transit across a discretized stellar disk tiled with realistic local spectra. In the present case, disk-integrated spectra are then convolved with the NIRPS instrumental response and resampled temporally within the windows of the observed exposures. Transmission spectra comparable to the observations are then computed in the same way by normalizing each simulated disk-integrated spectrum with the corresponding master stellar spectrum. We note that transmission spectra time series were computed for each transit independently and then fitted together to the entire dataset. 

\subsubsection{Stellar-spectrum modeling}\label{subsec:model_star}

EVE simulations require tiling a model stellar grid with realistic local spectra. Several effects such as the stellar rotation (which induces the Rossiter-McLaughlin effect; \citealt{rossiter_detection_1924,mclaughlin_results_1924}), center-to-limb variations \citep[CLV; e.g.,][]{vernazza_structure_1981,allende_prieto_center--limb_2004}, or limb darkening \citep[e.g.,][]{knutson_map_2007} can affect the spectral shape and position of the stellar lines across the stellar surface, so that the disk-integrated spectrum may be a poor proxy for the local spectra occulted by the planet along the transit chord.

\begin{figure}
\includegraphics[trim=0cm 0cm 0cm 0cm,clip=true,width=\columnwidth]{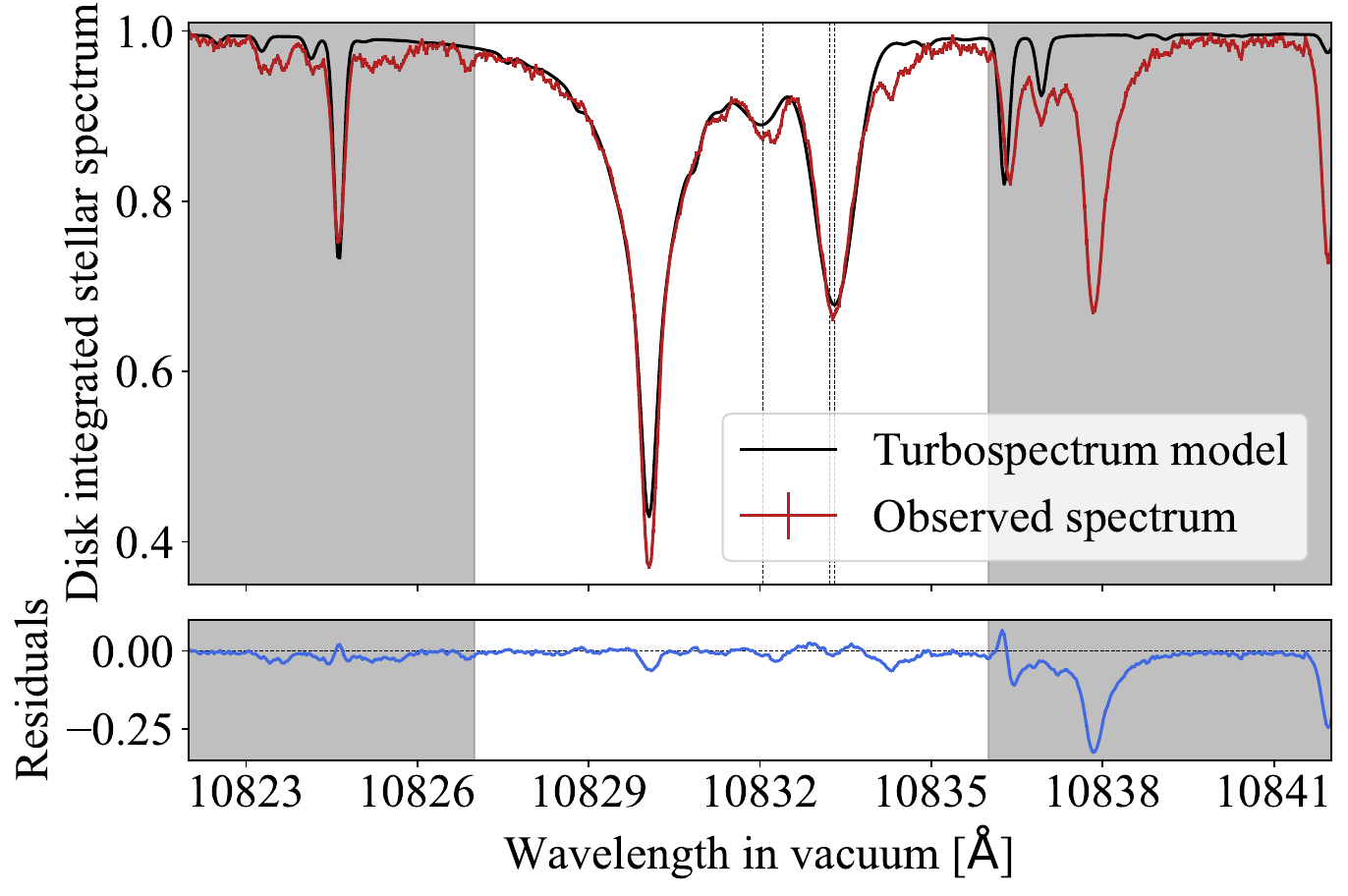}
\centering
\caption[]{Comparison between the modeled and observed out-of-transit disk-integrated stellar spectrum (\textit{top panel}) and residuals (\textit{bottom panel}). The gray areas represent regions excluded from the EVE fit to the transmission spectra (see Section~\ref{subsec:model_thermosphere}). Dashed vertical lines are the theoretical wavelengths of the helium triplet lines. The deep and broad line at $\sim 10830$~$\mathrm{\AA}$ corresponds to the aforementioned silicon line for which we fit the abundance.}
\label{fig:stellar_spectrum}
\end{figure}

Thus, we combined the 3 NIRPS visits to build a master out-of-transit spectrum (see Fig.~\ref{fig:stellar_spectrum}) that we fitted using the same 2D stellar grid as in the EVE code. This is a regular square-cell grid of the photosphere tiled with local spectra that account for the aforementioned stellar inhomogeneities. Local spectra are generated with the \textit{Turbospectrum} code\footnote{Latest version, available for download at \href{https://github.com/bertrandplez/Turbospectrum_NLTE/tree/master}{Turbospectrum$\_$NLTE}.} \citep{alvarez_near-infrared_1998,plez_turbospectrum_2012} and associated line lists (\citealt{heiter_atomic_2021,magg_observational_2022}; VALD3, \citealt{ryabchikova_major_2015}). Fixed parameters used to set the grid and generate the stellar models are summarized in Table\,\ref{table:params}. In particular, we note that knowledge of the stellar rotation is informed by the RM analysis (Sect.~\ref{sec:orb_arch}). We then adjusted the abundances of atomic species absorbing in the measured range by generating  corresponding series of local spectra, summing them over the model stellar grid, and comparing the resulting disk-integrated spectrum with the observed one \citep[a similar approach is used in][]{dethier_combined_2023}. 

Since the escaping helium atoms in the exosphere of \hbox{WASP-69\,b} have maximum radial velocities of a few tens of km$\cdot$s$^{-1}$, as revealed by the excess absorption in the blue wing of the measured signal, we focused our abundance fitting on lines at shorter wavelengths than the helium triplet, in particular the nearby, deep, and broad silicon line ($\sim 10830$~$\mathrm{\AA}$). 
We had to model the helium triplet spectrum independently as \textit{Turbospectrum} only reconstructs the photospheric spectrum of a star and does not include the chromosphere, from which the stellar helium triplet originates \citep{Vincenzo1997}. We used an analytical model\footnote{Original code from W. Dethier (priv. communication)} that computes the stellar helium triplet using Gaussian profiles for a given helium column density and temperature in the stellar chromosphere. The triplet is then multiplied directly into the local \textit{Turbospectrum} spectra. This is performed by shifting the helium triplet according to the local stellar surface rotational velocity derived from the RM analysis and adjusting the continuum level according to limb darkening. Our final stellar grid thus accounts for the stellar surface velocity field, limb darkening, and CLV, with the exception that CLV is not included for the chromospheric helium triplet.

Our best model for the disk-integrated spectrum reproduces the observed master-out well (Fig.~\ref{fig:stellar_spectrum}). Although we could not perfectly match the depth of the silicon line, highlighting limitations of current stellar atmospheric models, our best fit already provides a much better proxy for the local stellar spectrum than the measured disk-integrated master. We also note that a shallow line observed in the red wing of the helium triplet was not included in our fit. We did not find any correspondence in the line lists used by \textit{Turbospectrum} nor any evidence for telluric contamination, and thus speculate that this line either originates from a layer not included in the stellar atmospheric model or arises from a molecule. Since this feature is in the red wing of the triplet, it does not impact the fit of the absorption signal.

\begin{figure}
\includegraphics[trim=0cm 0cm 0cm 0cm,clip=true,width=\columnwidth]{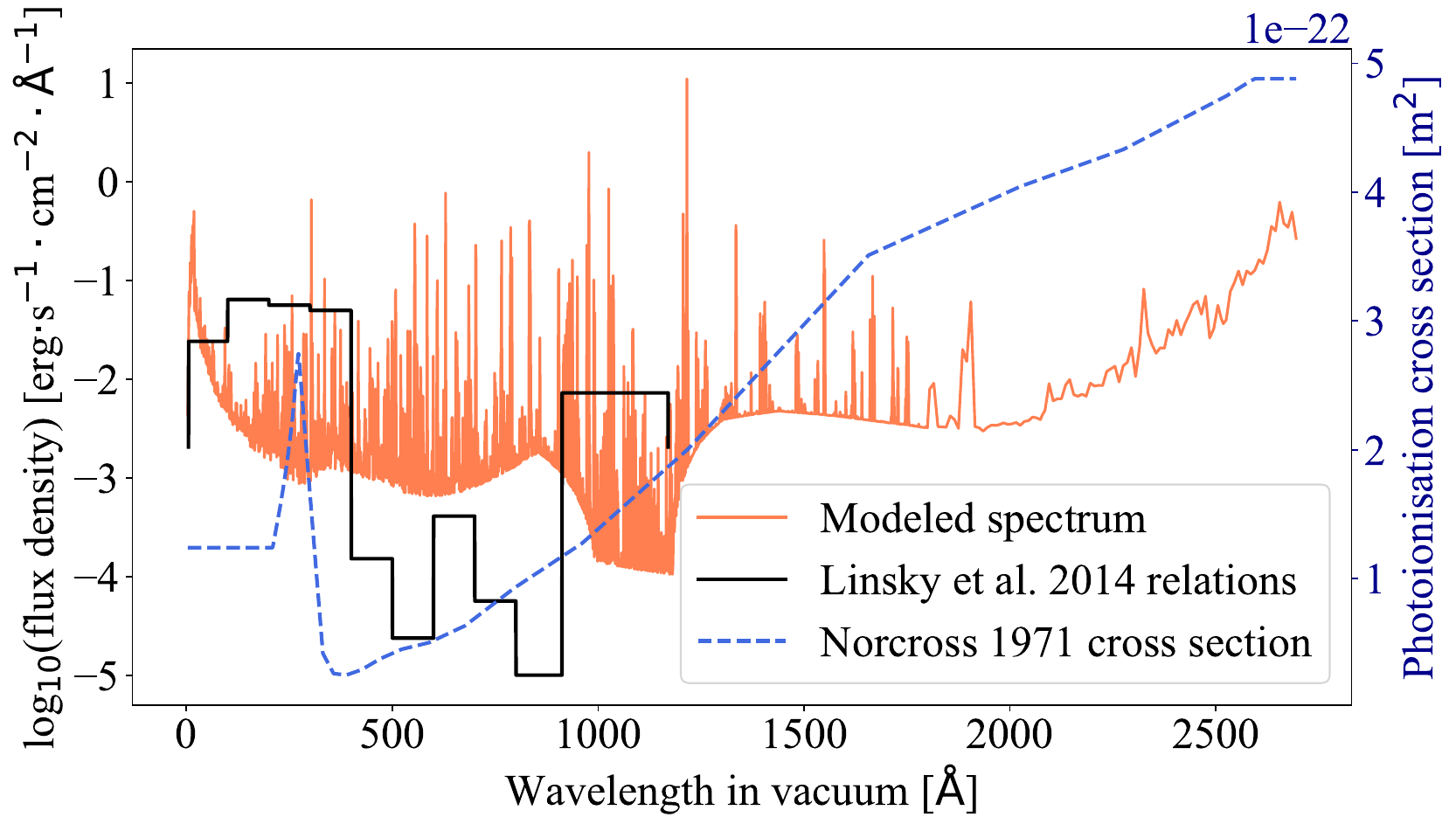}
\centering
\caption[]{Stellar XUV spectrum modeled (see Section~\ref{subsec:model_star}) and evaluated using \citet{linsky_intrinsic_2014} relations. The photoionization cross-section of the metastable helium triplet is overplotted. }
\label{fig:XUV_spectrum}
\end{figure}

The thermospheric chemistry and exospheric structure are highly dependent on the level of XUV irradiation received by the planet \citep[see e.g.,][]{oklopcic_helium_2019, gillet_self-consistent_2023}. The XUV flux is often estimated using the \citet{linsky_intrinsic_2014} scaling relations, providing rough estimates over broad spectral bands. Furthermore, these relations only estimate the XUV flux between a few $\mathrm{\AA}$ and 1200\,\AA. This should not limit the modeling of the thermospheric profiles of metastable helium since the population of this level is mainly controlled by the flux over the range $5-504$~$\mathrm{\AA}$ \citep{sanz-forcada_active_2008}. On the other hand, the photoionization cross-section of metastable helium atoms \citep{norcross_photoionization_1971} is large at wavelengths longer than $1200$~$\mathrm{\AA}$ (see Fig.~\ref{fig:XUV_spectrum}). This means that the photoionization of the outflow from the planet is underestimated when the XUV flux is not defined up to the helium ionization threshold.

To overcome this limitation, we built a stellar spectrum model that includes the usual XUV emission as well as the ultraviolet emission up to 2593~\AA. The emission in the XUV range arises from the coronal and transition region layers of the star, while the rest of the UV emission considered (~920--2593~\AA) also adds a photospheric contribution that becomes evident above $\sim$1800\,\AA\ (Fig.~\ref{fig:XUV_spectrum}). The photospheric contribution was estimated using the \citet{cas03} models with the stellar parameters from Table~\ref{table:params}, as detailed in \citet{lampon_characterisation_2023}. The coronal and transition region model, covering the $\log T$(K)$=4-7.4$ range, has been recently updated in \cite{Sanz_forcada_2025} using atomic data from ATOMDB 3.0.9 \citep{aped}, which improves the continuum emission at UV wavelengths. This model was made based on the analysis of the X-rays XMM-Newton observation of WASP-69 on 2016-10-21 (prop. ID 78356, PI: M. Salz, \citealt{Salz2015}). The XMM-Newton/EPIC spectra (exposure time 27.4~ks, fit in the range 0.4--3.5 keV) were fitted with a one-temperature plasma: $\log T$(K)$=6.73 \pm0.04$, emission measure $\log EM$(cm$^{-3}$)$=50.80\pm0.05$, $L_{\rm X}=(1.25\pm0.04) \times 10^{28}$ erg\,s$^{-1}$ (in 0.12--2.48~keV spectral range). The coronal model was then extrapolated to transition region temperatures following \citet{san11}, which is the main source of uncertainties in the final spectral energy distribution (SED). Finally, the synthetic spectrum produced with the coronal model was added to the photospheric contribution to produce a SED in the range 5--2593~\AA, as shown in Fig.~\ref{fig:XUV_spectrum}.

\subsubsection{Thermosphere modeling}\label{subsec:model_thermosphere}

The EVE code does not self-consistently model the thermospheric layer of the upper atmosphere. To generate metastable helium profiles we used the 1D \textit{p-winds} \citep{dos_santos_p-winds_2022} code. This code treats the highly irradiated expanding atmosphere as a Parker wind and computes the metastable helium density profile from the planetary parameters (Table\,\ref{table:params}) and the XUV flux received by the planet. Using the radiative transfer module included in \textit{p-winds}, we fitted the averaged absorption spectrum. We used this first-order approach to reduce the parameter space to be explored with EVE. 

Similarly to most helium signals, the in-transit absorption signature from \hbox{WASP-69\,b} is slightly blue-shifted (Sect.\,\ref{subsec:data_analysis}). This velocity shift is not correlated to the other parameters in the EVE fit to the absorption signal. Hence, we fixed the value we found, i.e., a blue-shift of $\sim 930$~m$\cdot$s$^{-1}$, from \textit{p-winds} in EVE. Moreover, we explored the impact of the theoretical H/He ratio and found in-transit absorption signals consistent with the data in an H/He range of $0.8$--$0.9$. We thus considered three different H/He ratios of 0.80, 0.85, and 0.90 in the fit performed with EVE.

\begin{figure}
\includegraphics[trim=0cm 0cm 0cm 0cm,clip=true,width=\columnwidth]{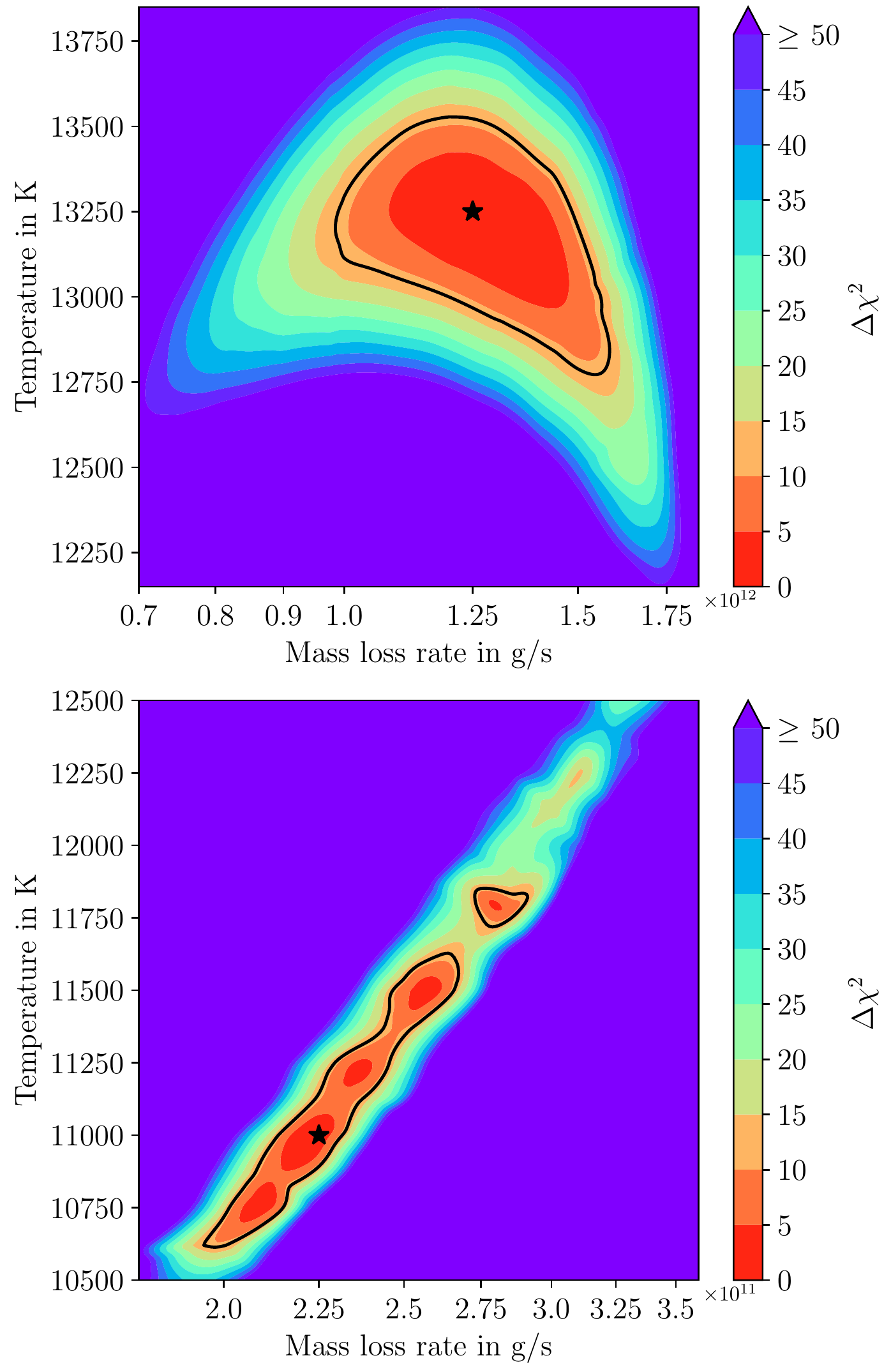}
\centering
\caption[]{$\chi^2$ maps of the EVE simulations. The maps are colored as a function of the difference between the projected minimum value of $\chi^2$ along the other parameters in the plane temperature-mass loss and the overall minimum $\chi^2$ corresponding to the best simulations (shown as black stars). The black line indicates the 3 $\sigma$ level in both panels. \textit{Top panel:} models with thermosphere only (see Section~\ref{subsec:model_thermosphere}), the H/He ratio varied between 0.8 and 0.9. \textit{Bottom panel:} models with thermosphere and exosphere (see Section~\ref{subsec:model_exosphere}), where the H/He ratio is fixed to 0.8.
}
\label{fig:maps_best_fit}
\end{figure}

In Fig.~\ref{fig:maps_best_fit}, we show the $\chi^2$ maps of the EVE transit models as a function of the temperature and the mass loss input in \textit{p-winds}. We explored a broad range of temperatures and mass loss rates, but the high S/N of the three combined transits gives a very good constraint on these two parameters. The best fit is obtained for $T=13\,210^{+99}_{-108}$ K and $\dot{M}=(1.25^{+0.09}_{-0.07})\cdot 10^{12}$~g$\cdot$s$^{-1}$ for a H/He ratio of 0.80. 

We also show in Fig.~\ref{fig:TS} and Fig.~\ref{fig:LC}, respectively, the averaged in-transit absorption spectrum and the helium light curve for the best-fit model. We see that the thermosphere alone is not able to reproduce all the features in the helium absorption profile. Indeed, while the central part of the doublet is well-fitted, this is not the case for its blue wing and the third component. This is likely due to the velocity field of the escaping tail (which is not accounted for in the thermospheric simulations), which would produce blueward absorption of the helium transitions. Moreover, the asymmetry of the NIRPS light curve, especially the post-transit absorption, cannot be reproduced by a model of a spherical thermosphere, once again indicating that an exosphere must be included.

\subsubsection{Exosphere modeling}\label{subsec:model_exosphere}

In the previous section, we saw that the NIRPS observations cannot be explained by a thermospheric contribution alone. This motivated us to include escaping particles as another source of absorption. Similarly to the previous section, we use \textit{p-winds} to generate a thermospheric structure that we couple with the exospheric contribution simulated by EVE using a Monte Carlo particle description. These particles are then subject to gravity, radiation pressure, and photoionization. EVE derives a metastable helium mass loss from the thermospheric density profile and the vertical velocity. It also recomputes an effective mass loss to account for particles that fail to escape the planet's gravitational well and eventually fall back within its thermosphere. 

As a first approach, we explored regions of the parameter space close to the best fit obtained without the exosphere. We found that no metastable helium cometary tail forms with the nominal spectrum of the star (XUV and bolometric), as atoms are photoionized instantaneously after escape (the photoionization lifetime is just $\sim$4~minutes in this case). This is very similar to \hbox{WASP-107\,b} \citep[see][]{allart_high-resolution_2019}, for which reproducing the observed absorption with a simulated exosphere required decreasing both the bolometric and XUV stellar fluxes. We thus scaled down the flux of WASP-69 similarly by a common factor in both energy ranges. We explored a wide range of decreasing factors and found a best value of around 200, which we fixed for the rest of the exploration. We note that this reduced stellar flux is only used to compute the photoionization of the exosphere, as we still used the nominal stellar flux to generate the thermospheric structure. Generating a cometary tail of metastable helium indeed also requires sufficient mass loss from the thermosphere, which would not be possible with such a reduced XUV irradiation. This suggests that our reconstruction of the stellar XUV spectrum is correct and that the effective flux received by helium atoms in the exosphere is reduced. We discuss possible scenarios in Sect.~\ref{subsec:comparison_model}.

Because we needed to introduce such a strong assumption on the XUV and bolometric fluxes, we decided to also fix the H/He ratio to 0.80 (the value favored by the fit of the thermosphere alone) to simplify the analysis. Furthermore, we varied the value of the exobase between 1.5 planetary radii and the Roche Lobe (2.92 R$_{\mathrm{p}}$) and found that the best fits are always obtained for the Roche Lobe. Yet, with this approach, we were never able to reproduce the absorption level that we see in the second part of the light curve. In our data, the long post-transit absorption indicates that the escaping matter goes far away from the planet. However, the asymmetry of the absorption profile in the blue wing is moderate, suggesting a relatively slow velocity field for the exosphere. This is surprising because such a slow velocity field cannot move enough metastable helium atoms far enough from the planet in a few hours (limited by the photoionization and de-excitation lifetime) to explain the observed post-transit absorption. This is why we propose that the NIRPS data hints at a repopulation of the metastable helium level inside the tail and that a fluid model might be able to reproduce the post-transit features more accurately. 

Firstly, we tried to mimic this effect by simulating exospheres without the natural de-excitation of the metastable helium level so that the tail is only depopulated through photoionization. The results of these simulations are shown and discussed in Appendix\,\ref{app:escape}. In this case, the exosphere was very puffy and dense around the planet, and it has a similar effect to increasing the thermospheric radius. We were able to reproduce the long post-transit absorption, as well as the absorption spectrum after the transit was, suggesting the presence of a tail beyond the planet. This tail must be sufficiently dense to produce about 1\% of absorption (see Appendix\,\ref{app:escape}). This supports the idea that a low H/He is favored, as it results in an atmosphere with a higher mass-loss rate of metastable helium. However, it is challenging to distinguish between the effect of this ratio, mass-loss rate, and temperature without additional observations of escaping hydrogen. On the other hand, the average absorption spectrum during the transit was on the right order of magnitude, but highly blue-shifted due to the high velocity reached by the particles. We thought that a fluid model would restrain the velocity of the metastable helium and might match the weak blue shift of the data. We thus decided to explore thermospheres more extended than the Roche Lobe. This naturally allows us to also vary the geometry of the thermosphere (corresponding to the fluid regime in our simulation) and no longer restrict ourselves to spherical atmospheres, we discuss the physical motivation of this in Sect.~\ref{subsec:comparison_model}. 

\begin{figure}
\includegraphics[trim=0cm 0cm 0cm 0cm,clip=true,width=\columnwidth]{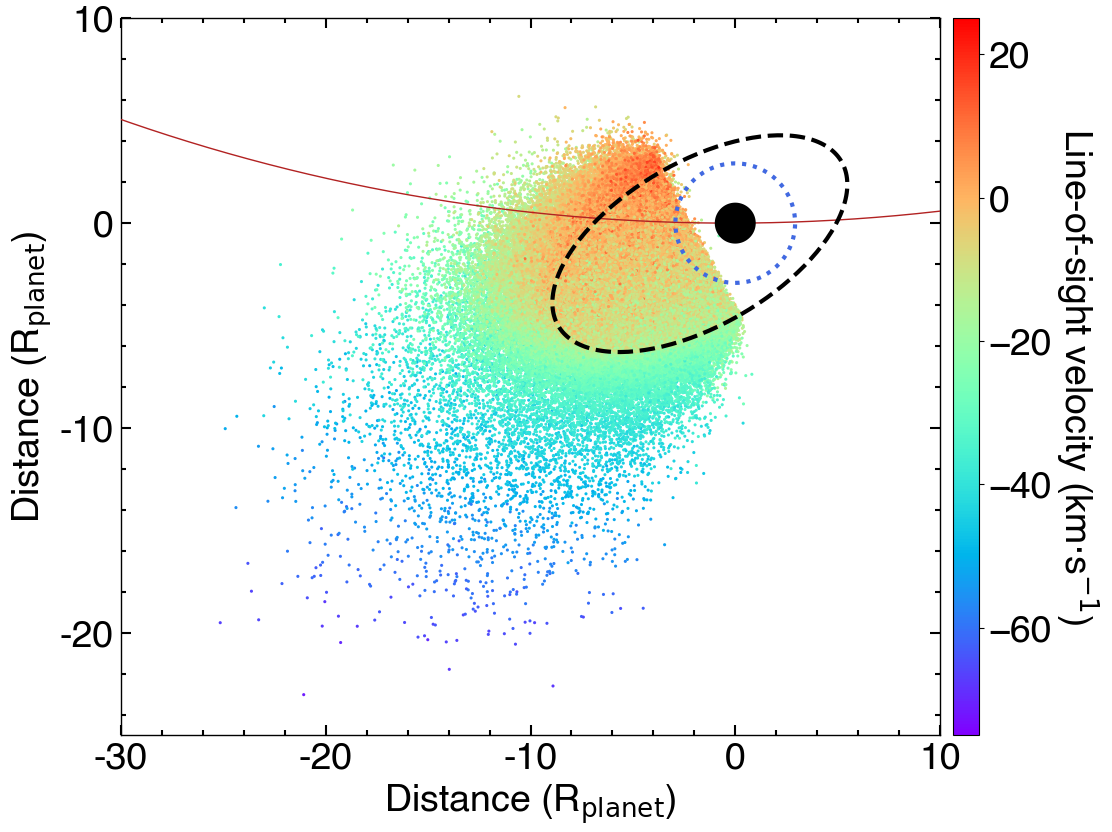}
\centering
\caption[]{View of the system in the best-fit simulation, including an exosphere, as seen from the perpendicular to the orbital plane at mid-transit. The color bar indicates the line of sight velocity towards the observer in km$\cdot$s$^{-1}$. The black dashed line corresponds to the exobase and the blue dashed line to the Roche lobe. The cometary tail extends out to 17 times the planetary radius. }
\label{fig:above}
\end{figure}

We explored a couple of geometrical configurations without necessarily covering the full parameter space (which is beyond the scope of this paper). The best fits were obtained for elliptical thermospheres that extend up to 10 R$_{\mathrm{p}}$ after the solid core of the planet (see Fig.~\ref{fig:above}). Figure~\ref{fig:maps_best_fit} displays the $\chi^2$ map of EVE simulations as a function of the thermospheric temperature and mass loss rate. We see that the parameter space is better constrained when accounting for the exosphere. The derived mass loss is lower when the exosphere is included. Indeed, escaping material now contributes to the absorption and the thermosphere itself is more extended so that the local contribution of the thermosphere to the absorption must be smaller, and its density of helium consequently lower. This is why the derived mass loss is lower, as it controls the thermospheric density in the model. We find the best fit for $T=11\,000$ K and $\dot{M}=2.25\cdot 10^{11}$~g$\cdot$s$^{-1}$.

\begin{figure}
\includegraphics[trim=0cm 0cm 0cm 0cm,clip=true,width=\columnwidth]{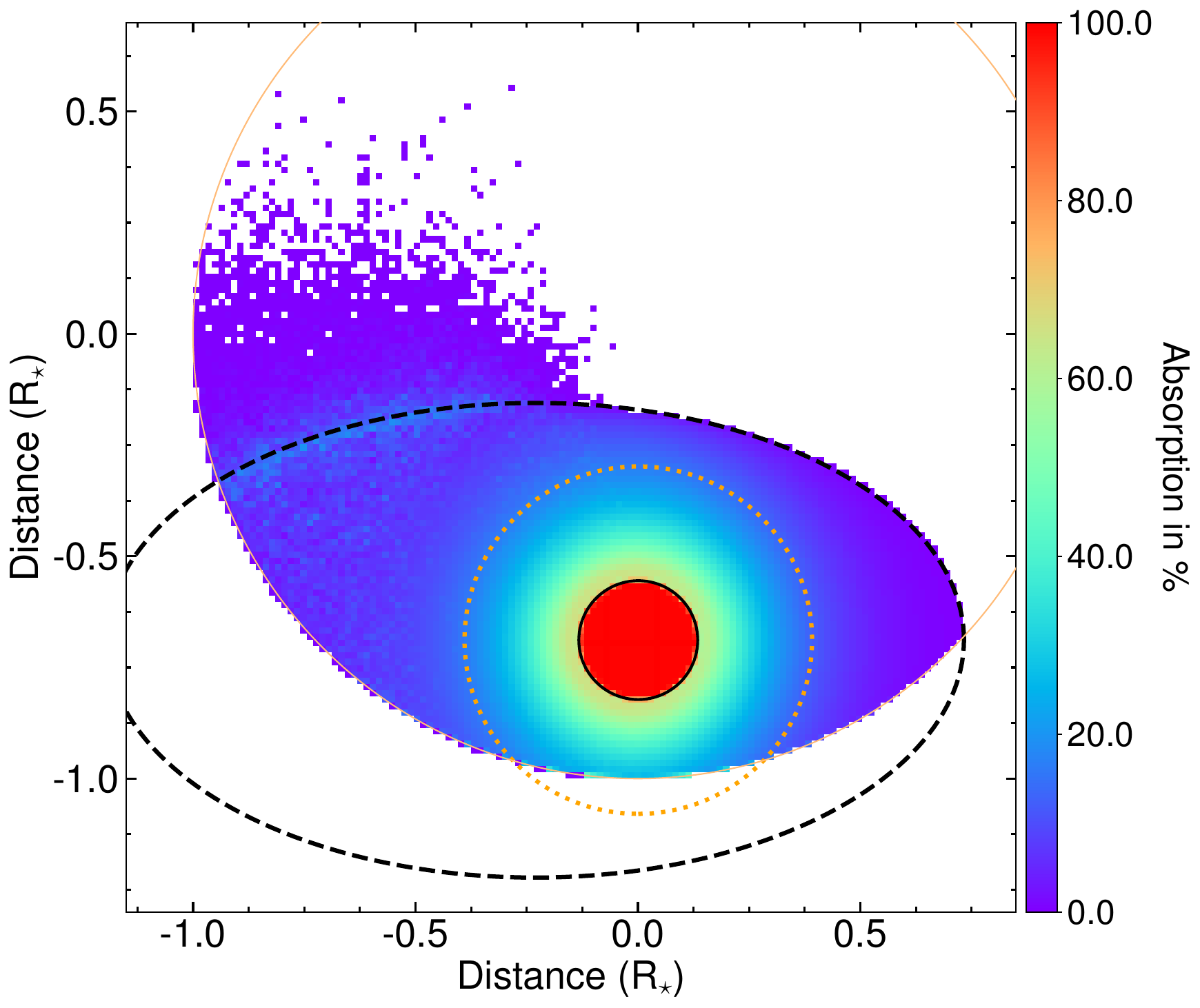}
\centering
\caption[]{Line of sight absorption of the transiting planet in front of the stellar disk at mid-transit. The absorption level indicates the amount of flux from the local stellar cell that is absorbed. The black dashed line corresponds to the exobase and the orange dashed line to the Roche lobe. Outside the stellar surface, there is no flux occultation so no absorption is seen. }
\label{fig:camera}
\end{figure}

With the presence of an exosphere, the simulated in-transit absorption profile (Fig.~\ref{fig:TS}) is broader, and the transition between the doublet and third line of the triplet is smoother. 
The metastable helium atoms escaping the planet acquire a radial velocity gradient towards the observer due to stellar radiation pressure (see Fig.~\ref{fig:above}), causing absorption at various blue shifts and resulting in a smooth transition in the blue wings of all three lines.
Typical velocities inside the cometary tail are on the order of a few tens of km$\cdot$s$^{-1}$. 
We note that while the averaged absorption spectrum with an exosphere captures the main features seen in the data, it does not fit them perfectly, especially for the third line of the triplet.
The simulated light curve associated with the best fit is asymmetric and fits the observed light curve well (see Fig.~\ref{fig:LC}). The simulated excess absorption follows the observation very well, except at the very end of the absorption, where it is slightly overestimated. This indicates that our best-fit cometary tail, which extends out to about 17 times the planetary radius (see Fig.~\ref{fig:above}), might be a little bit too dense at larger distances from the planet, but it remains within the error bars. In Fig.~\ref{fig:camera}, we show a view of the local absorption caused by the transiting planet on the stellar disk. The opaque layer, presenting complete absorption, is clearly visible, with the thermosphere surrounding it and symmetrically absorbing up to $\sim40\%$ where the metastable density profile peaks. Further, from the planet's center, the puffy exosphere generates weaker absorption, trailing the planet. This is due to the launching of the particles that are performed at the end of the cometary tail (discussed in Sect.~\ref{subsec:comparison_model}). We also notice that the exosphere absorbs above the planet center in the stellar disk projected view. This comes from the assumption that the outflow is isotropic in this simulation.



\section{Discussion of the helium signature}\label{sec:discussion}
\subsection{Atmospheric dynamics}
The helium signature of WASP-69\,b is well detected in the three transits observed with NIRPS leading to exquisite precision on the atmospheric dynamics thanks to  high temporal sampling and spectral resolution (Fig.\,\ref{fig:TS_param}). We observe a velocity field of the upper atmosphere going from 4.74$\pm$2.93\,km$\cdot$s$^{-1}$ to $-$29.46$\pm$2.46\,km$\cdot$s$^{-1}$ with an average line width of $\sim$43\,km$\cdot$s$^{-1}$. 
However, these velocities are measured in the line of sight and do not consider the impact of planetary rotation. Assuming that the planet is tidally locked, we can estimate the planetary rotation to be $\sim$1.4\,km$\cdot$s$^{-1}$ leading to a rotation of 4.3\,km$\cdot$s$^{-1}$ at the top of the atmosphere probed by the helium triplet (2.3\,R$_p$). From there, the contribution in the line of sight at the equator is about two-thirds, leading to an increase of the velocity field of 2.9\,km$\cdot$s$^{-1}$. Therefore, the velocity field is more likely to be non-redshifted at egress to still highly blueshifted after transit. This could be explained by a combination of day-to-night side winds in the thermosphere \citep{allart_spectrally_2018} and material blown away under the impact of stellar irradiation in the exosphere \citep{allart_high-resolution_2019}.
In a similar way, the line width velocity corresponds to the maximum wind speed from which the line of sight contribution has to be removed, leading to a broadening of $\sim$28.9\,km$\cdot$s$^{-1}$. This could stem from either vertical, homogeneous winds or a global super-rotational wind \citep{seidel_wind_2020, seidel_high_2023}. An in-depth analysis is beyond the scope of this paper.

\subsection{Comparison with the literature}
\begin{figure}
\includegraphics[width=\columnwidth]{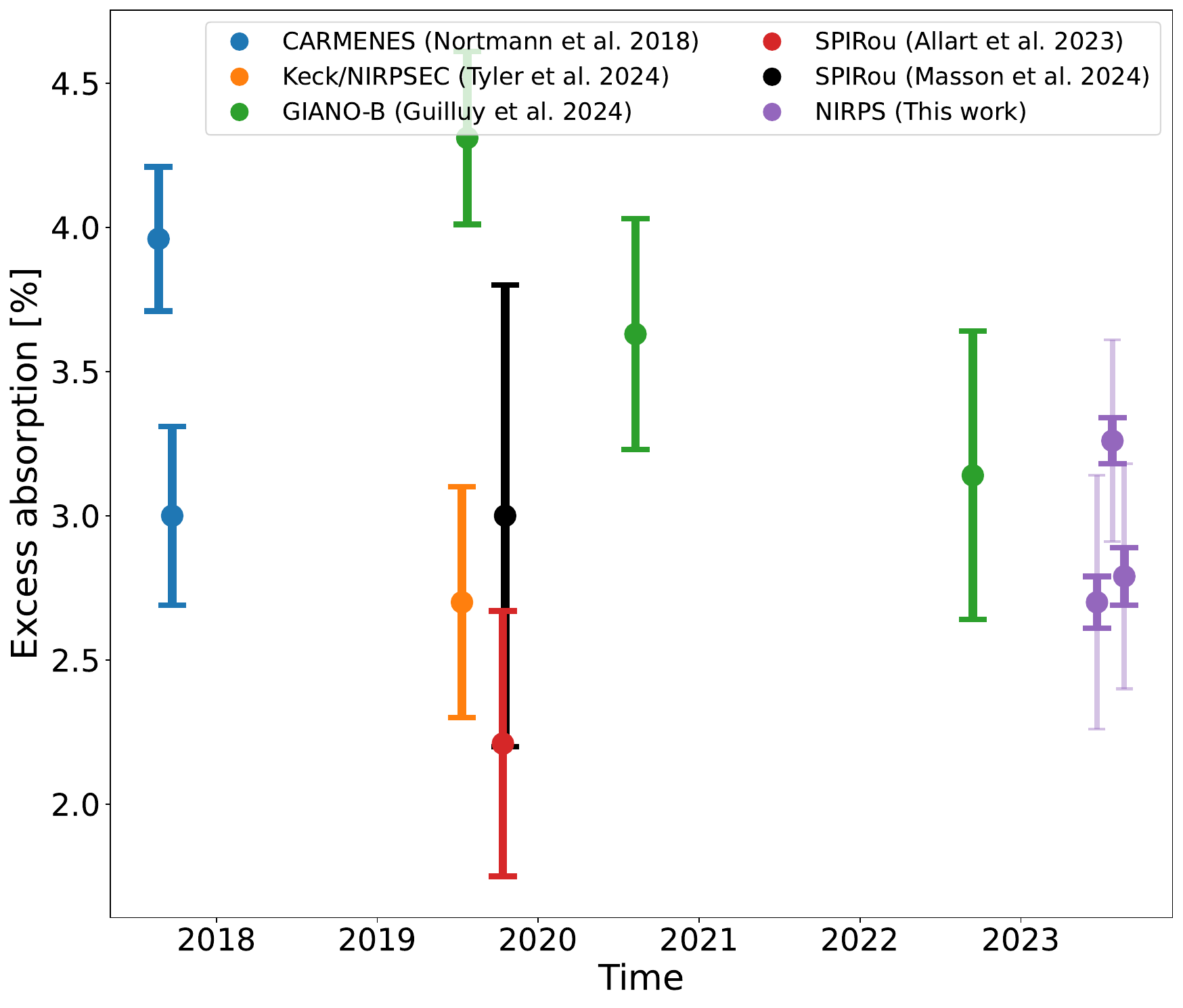}
\centering
\caption[]{Excess absorption of helium as a function of time reported by multiple teams using high-resolution instruments: CARMENES \citep{nortmann_ground-based_2018}, Keck/NIRSPEC \citep{tyler_wasp-69bs_2024}, GIANO-B \citep{guilluy_gaps_2024}, and SPIRou \citep{allart_homogeneous_2023,masson_probing_2024}. The SPIRou analysis of \cite{masson_probing_2024} was offset by five days compared to \cite{allart_homogeneous_2023} for visualization purposes. In light purple are the 1-$\sigma$ dispersions of the continuum measured on the NIRPS transits. This is to allow a better comparison with the values in the literature.}
\label{fig:lit_comp}
\end{figure}

The helium signature strength is in good overall agreement with previously published results \citep{nortmann_ground-based_2018,tyler_wasp-69bs_2024,guilluy_gaps_2024,allart_homogeneous_2023, masson_probing_2024} as presented in Fig.\,\ref{fig:lit_comp}. However, we want to point out that discrepancies on how the excess absorption and its uncertainty\footnote{In addition, the data reduction and computation of the transmission spectrum differ slightly from one instrument and analysis to another.} are reported in the literature make the interpretation of signal variability between transits more difficult. We thus provide in Appendix\,\ref{app:excess_computation} the excess absorption measured with different techniques for our NIRPS observations.\\
Overall, our three transits are in good agreement, but we observe some variability, particularly in the red wing of the helium triplet. While pinpointing the origin of such differences remains difficult, we can confidently exclude both telluric residuals and the data extraction\footnote{Similar behavior is observed with the NIRPS-DRS pipeline and with APERO}, making potential stellar or planetary variability more likely explanations. Studying the intra-transit variability of escaping atmospheres is difficult, as multiple factors can impact the observations, but it would provide significant inputs to our understanding of star-planet interaction and the evolution of exoplanets. Such precise studies require the homogeneous analysis of datasets using the same extraction and reduction pipelines and, ideally, the same instrument. We aim to provide such analysis by continuing to observe WASP-69\,b during the whole NIRPS GTO duration. \\

We also confirm that the helium signature is blueshifted during transit, which is in agreement with the CARMENES, GIANO-B, and Keck/NIRSPEC observations \citep{nortmann_ground-based_2018,guilluy_gaps_2024,tyler_wasp-69bs_2024}. Our fine temporal sampling of the velocity shift of the signature allows us to distinguish four different regimes: before and during ingress, the signature is redshifted; between t$_2$ and mid-transit, the absorption is slightly blue-shifted; between mid-transit and t$_4$, the metastable helium particles move toward us ($\sim$$-$6\,km$\cdot$s$^{-1}$); after egress, the particles are accelerated away from the star at even higher velocities, up to $\sim$$-$30\,km$\cdot$s$^{-1}$. The post-transit absorption lasts for approximately 50 minutes after egress, which is similar to GIANO-B \citep{guilluy_gaps_2024}, but shorter than estimated with Keck/NIRSPEC (1.28h, \citealt{tyler_wasp-69bs_2024}) and longer than with CARMENES (22min, \citealt{nortmann_ground-based_2018}). The discrepancy with the latest study might result from its shorter post-transit baseline or due to how the helium light curve was built. The difference with Keck/NIRSPEC could reflect instrumental systematics and, in particular, spectrograph stability that might have biased the analysis of the single transit presented in \cite{tyler_wasp-69bs_2024}. \\
\cite{levine_exoplanet_2024} reported a lower absorption for the transit occurring on the night of 2023-08-24, corresponding to our third transit. However, we do not detect lower excess absorption from this transit than other transits observed with NIRPS or other high-resolution instruments. Therefore, we tentatively attribute this reduced absorption to the instrumental systematics of Palomar/WIRC. \\

\subsection{Helium signature modeling interpretation and comparison with the literature}\label{subsec:comparison_model}

Together with the large number of observations of \hbox{WASP-69\,b}, the literature contains a variety of mass loss and thermospheric temperature estimates. 
We recall that this analysis yields a temperature of $11\,000$~K and establishes a mass loss rate of $\dot{M}\sim 2.25\cdot 10^{11}$~g$\cdot$s$^{-1}$.
The simplest models used to derive these parameters are Parker wind 1D codes. \citet{allart_homogeneous_2023} found mass loss value consistent with ours within \hbox{3 $\sigma$}. \citet{Vissapragada2022} reported a strong degeneracy between the mass loss and the temperature when fitting their Palomar/WIRC observations. Our best-fit temperature of $T\sim11\,000$\,K is associated with a mass loss slightly larger than $\dot{M}\sim 10^{11}$~g$\cdot$s$^{-1}$ (against $\dot{M}\sim 2.25\cdot 10^{11}$~g$\cdot$s$^{-1}$ in our study), assuming a H/He ratio of 0.90. The mass-loss rate we find is almost at the edge of their energy limited contour. Using a Parker wind model coupled with a more complex chemistry that includes metals to fit the Palomar/WIRC dataset, \citet{Linssen_2022} found mass losses reduced by a factor of 10 and temperature dropping to $T\sim5000$ K. Other studies used 1D atmospheric codes solving for hydrodynamics equation. Using this approach, \citet{lampon_characterisation_2023} found a much lower temperature than we do ($T\sim5800$ K) and $\dot{M}\sim 10^{11}$~g$\cdot$s$^{-1}$ while \citet{Caldiroli_2022} found $\dot{M}\sim 5 \cdot 10^{11}$~g$\cdot$s$^{-1}$. Finally, \citet{wang_metastable_2021} used a 3D hydrodynamical code and reported $\dot{M}\sim 3\cdot 10^{11}$~g$\cdot$s$^{-1}$, but had no evidence for a cometary tail in their data. This illustrates the wide variety of tools available in the community to derive mass loss rates of evaporating exoplanets. Overall, all studies to date required strong assumptions but agreed with a mass loss rate of the order $10^{11}$~g$\cdot$s$^{-1}$ for WASP-69\,b. We note that most of previous studies used a rather simple proxy for the XUV flux, similar to what is shown in Fig.~\ref{fig:XUV_spectrum} for the \citet{linsky_intrinsic_2014} relations. They also frequently used a H/He ratio of 0.90 or higher, while we found that our data are more consistent with a lower H/He ratio (0.80 in our analysis). Finally, they assumed spherical geometry for the thermosphere.
This could, in part, explain differences with our derived mass loss. It emphasizes once again the major role of the XUV spectrum in driving the metastable chemistry as well as the assumption made over the H/He ratio, two quantities that are very poorly constrained. We further note that we fixed the H/He ratio to 0.8 in our analysis. In practice, variations in the H/He ratio could influence the mass-loss rate determined in this study. Future observations of hydrogen escape from this planet could help resolve this degeneracy.

Regarding our tail modeling, we attribute the decrease of the XUV and bolometric fluxes to self-shielding within the tail or absorption by an unknown absorber between the planet and the star \citep[similar to][]{allart_high-resolution_2019}. However, we note that \citet{levine_exoplanet_2024} reported a decrease in XUV flux by at least a factor of 2 since 2016. As our XUV model is constrained using observations from that time, this could partly account for the discrepancy.
Another key point of this analysis is our assumption of a non-collisional regime above the exobase. As a result, there is no mechanism to repopulate the metastable level inside the tail, and helium atoms can only de-excite or be photoionized. This approach is completely different from hydrodynamical models in which planetary outflows are assumed to remain collisional even at a large distance from the planet. For instance, \citet{MacLeod_2022} simulated leading and trailing tails of matter from the interaction between the planetary outflow and the stellar wind. In this model, the escaping matter is considered as a fluid, and thus, the metastable helium level is repopulated through the interaction with the stellar wind itself. This model highly depends on the assumed stellar wind configuration but is able to reproduce pre- and post-transit helium light curves. The overall shape of our thermosphere is not so far from what is predicted by such a fluid model with moderate stellar winds. Indeed, usually, the flow of matter is compressed and elongated in a direction that is slightly tilted compared to the orbit's tangent. Note that we tried several inclinations with respect to the orbit's tangent but did not fit the value in this analysis. We also launched particles in a non-collision regime only from the back of the thermosphere to be more consistent with a flow shaped by the stellar wind interaction. The outflow is, in fact, not expected to escape at the cometary tail's front because its velocity reverses due to the stellar wind confinement.

Finally, we compared the mass loss we found to the energy-limited value given by \cite{Erkaev2007} and \cite{Caldiroli_2022}
\begin{equation}
    \dot{M}= \eta_\mathrm{eff} \frac{3 \phi_{\mathrm{XUV}}}{4 K G \rho_{\mathrm{pl}}}.
\end{equation}
In this equation, $K$ is a correction factor to account for tidal contribution, $\phi_{\mathrm{XUV}}$ is the integrated XUV flux between 5 and 912~\AA, and $\rho_{\mathrm{pl}}$ is the planet bulk density. To estimate the maximum mass loss driven by photoionization in a very conservative way, we assumed the highest efficiency for XUV heating $\eta_\mathrm{eff}$. Using \citet{Caldiroli_2022} framework we found an outflow efficiency of $\eta_\mathrm{eff} \sim 0.6$ which ultimately translates to a maximum mass loss of $1.90\cdot 10^{11}$~g$\cdot$s$^{-1}$. This is slightly smaller than the value we derived from the data but reasonably close. This estimate is given for the sub-stellar point (due to 1D modeling); however, planets are not uniformly irradiated. As such, observational estimations might lead to more realistic constraints; for instance, \cite{KrishnamurthyCowan2024} estimated a value of $6\%$ for planets with helium detection and predicted a maximum mass loss of $\sim$ 6$\cdot$10$^{10}$~g$\cdot$s$^{-1}$.

All of these points suggest that the mass-loss rate we derived from our simulations is relatively robust. The data hint at an interaction between the stellar winds and the planet's upper atmosphere. As a result, the fluid regime of this atmosphere is in a more exotic configuration than expected. Previous studies agree with our mass loss in general, as does the Energy-limited formula. Our analysis highlights that simple Parker wind models, when combined with non-collisional escaping particles, are not adequate to reproduce this specific observed signal and overestimate the planet's mass loss if the shape of the tail is not well modeled.

\subsection{Atmospheric escape in the context of a misaligned orbit}
With its relatively young age (1.3 $\pm$ 0.4\,Gyr), its slightly misaligned orbit ($\psi$ = 29.2$^{+6.1}_{-5.0}$ $^{\circ}$) and its escaping atmosphere ($\dot{M}=2.25\cdot 10^{11}$~g$\cdot$s$^{-1}$), \hbox{WASP-69\,b} joins an increasing number of giant planets (such as GJ\,436\,b \citep{ehrenreich_giant_2015,bourrier_orbital_2018}, GJ\,3470\,b \citep{bourrier_hubble_2018,stefansson_polar_2022}, HAT-P-11\,b \citep{allart_spectrally_2018,Ben-Jaffel_Neptune_2022,bourrier_dream_2023} and WASP-107\,b \citep{allart_high-resolution_2019,Rubenzahl_retro_2021}) at the edge of the Neptunian desert which share these peculiar properties. Assuming that the mass-loss rate of \hbox{WASP-69\,b} remains constant in the future, we calculated that the planet will lose up to 14\% of its mass over the next 10\,Gyr, meaning that this planet is very stable at the upper edge of the Neptunian desert \citep[similarly to][]{Vissapragada2022}. Had these planets migrated early on within their protoplanetary disks, they would have kept the primordial alignment or near alignment of their orbital planes as their stars would have tidally circularised their orbits, and they may have lost their atmospheres as they entered the desert. Thus, it has been proposed that a fraction of Neptunes arrived at their close-in orbits through high-eccentricity migration induced by an outer companion \citep{bourrier_orbital_2018,owen_photoevaporation_2018,Hang_high_2024}, for example through the Kozai-Lidov process \citep{kozai_secular_1962,fabrycky_shrinking_2007}. This type of migration naturally leads to a highly misaligned orbit because it can occur hundreds of millions of years to billions of years after the formation of the planet. It would explain how the orbit of these late Neptunian migrators has not yet circularized and how they survived atmospheric erosion as it started recently and is weaker due to the lower irradiation from their older star. These two possible pathways for the origin of close-in Neptunes may underlie the dichotomy between the Neptunian desert and savanna  \citep{bourrier_dream_2023}. Moreover, it underlines the importance of studying their atmospheric and dynamical evolution together \citep{attia_jade_2021} through observational constraints on the orbital architecture and mass loss obtained. Spectrographs like HARPS and NIRPS are ideal for this, and the NIRPS consortium has included such studies in its objectives.

\section{Conclusions}\label{sec:conclusion}
In this paper, we present the first results from NIRPS related to in-depth spectral characterization of exoplanets. We combined for the first time the analysis of the orbital architecture with the characterization of the upper atmosphere through the near-infrared helium triplet for WASP-69\,b. Our results reveal that the atmosphere of WASP-69\,b extends well above the Roche Lobe radius (3.17$\pm$0.05\% over the 0.75\,\AA\ passband, equivalent to 2.3\,R$_p$) and that metastable helium particles are escaping at a rate of $2.25\cdot 10^{11}$~g$\cdot$s$^{-1}$, forming a cometary-like tail observable up to 50 minutes after the end of the planet's optical transit. The spectral signature is well resolved spectrally and temporally, thanks to the high quality of the NIRPS data and the absence of systematics, which leads to an unprecedented characterization of the atmospheric dynamics. The helium velocity field is accelerating along the transit chord. It is maximal 50 minutes after the end of the transit when most of the helium particles are blown away toward us by the stellar irradiation. To best model the complex shape of the helium signature, we used the 3D code EVE \citep{bourrier_3d_2013} to simulate for the first time both the thermosphere and exosphere of WASP-69\,b. Our best-fit model favors an elliptical thermosphere with a H/He ratio of 80\,\% at high temperatures (11\,000\,K), which extends up to 10 R$_\mathrm{pl}$ and is impacted by slow day-to-night side winds. The exosphere models the velocity field observed with NIRPS in the form of a cometary tail with a length of up to 17 times the planetary radius with decreasing density. With this large exosphere, the best-fit model can reproduce the absorption observed post-transit and suggests that chemical processes repopulating the tail from within are necessary as well as interaction with the stellar wind (e.g. fluid models, \citealt{MacLeod_2022}). \\
The three NIRPS transits show moderate variability in the helium profile compared to the variability reported in the literature for WASP-69\,b \citep{nortmann_ground-based_2018,guilluy_gaps_2024}. We can exclude telluric and instrumental systematics with confidence, letting pseudo-stellar and planetary variability be the main culprit. While pseudo-stellar variability can impact the absorption strength, it cannot reproduce all the variability we measured. In addition, we do not observe simultaneous spot-crossing events in photometry. Therefore, we are left with planetary variability to explain the difference in the helium line shape between transits. A large-scale reanalysis of all available transits of WASP-69\,b with a homogeneous reduction is required but beyond the scope of this paper.\\
To the best of our knowledge, this is the first time that an RM analysis has been performed homogeneously over nine independent transit datasets, using a state-of-the-art pipeline, \textsc{antaress} (\citealt{Bourrier2024}). The very good agreement between the spin-orbit angle of WASP-69b derived from NIRPS and HARPS/HARPS-N data strengthens our result and highlights the consistency of RM signals between the near-infrared and optical domains. Furthermore, we search for long term variability of the projected spin-orbit angle but we do not observe evidence of stellar precession over a decade of observations. NIRPS will thus play an essential role in measuring the orbital architectures for planets around cooler stars, too faint at optical wavelengths. Our final measurements for $\lambda$ = 0.05$\pm$1.10$^{\circ}$ and $\psi$ = 29.2$^{+6.1\,\circ}_{-5.0}$ show that the orbital plane of WASP-69\,b is actually slightly misaligned and highlight the observational bias shown by \cite{attia_dream_2023} between the projected and 3D spin-orbit angles. If a high-eccentricity migration scenario is retained to best explain the mass loss rate and misalignment measured with the exquisite NIRPS and HARPS observations, then astrometry from Gaia could reveal the presence of an external companion. Moreover, long-term RV monitoring could provide better constraints on a remaining eccentricity.\\
From our results obtained with NIRPS, we can say that among the scarce population of close-in intermediate-mass planets, WASP-69\,b joins an increasing fraction of planets known for actively losing mass and having misaligned, often eccentric, orbits \citep[e.g.][]{correia_why_2020,bourrier_dream_2023}. Combining these properties with the expected diversity of composition for intermediate-mass planets \citep{moses_compositional_2013}, they form a unique sub-population to better understand how planets form and evolve. Taking the mass-loss rate we derived, WASP-69\,b will lose up to 14\% of its mass in the next 10\,Gyr. It indicates that the planet will remain at the upper edge of the Neptunian desert.

The instrumental stability and excellent performances of NIRPS, an AO-assisted high-resolution near-infrared spectrograph, allow us to push further the study of exoplanet atmospheres and orbital architectures than its competitors on similar-sized telescopes. Combined with the largest observational program (225 nights) secured over five years for such studies by the NIRPS consortium, it will allow us to lead unparalleled studies from the in-depth analysis of high-S/N, high-fidelity datasets to population surveys. Fundamental questions on atmospheric properties (escape, composition, chemistry, dynamics, variability, ...) will be answered and will improve our understanding of the formation and evolution of exoplanets.

\begin{acknowledgements}
We thank the anonymous referee for their fruitful comments that help improved the overall quality of the paper.
This work has made use of the VALD database, operated at Uppsala University, the Institute of Astronomy RAS in Moscow, and the University of Vienna.\\
This work presents results from the European Space Agency (ESA) space mission Gaia. Gaia data are being processed by the Gaia Data Processing and Analysis Consortium (DPAC). Funding for the DPAC is provided by national institutions, in particular the institutions participating in the Gaia MultiLateral Agreement (MLA). The Gaia mission website is \url{https://www.cosmos.esa.int/gaia}. The Gaia archive website is \url{https://archives.esac.esa.int/gaia}.\\
RA acknowledges the Swiss National Science Foundation (SNSF) support under the Post-Doc Mobility grant P500PT\_222212 and the support of the Institut Trottier de Recherche sur les Exoplan\`etes (IREx).\\
RA, FBa, CC, \'EA, BB, NJC, RD, LMa, LB, AB, AD-B, LD, AL, OL, LMo, JS-A, PV, TV \& JPW acknowledge the financial support of the FRQ-NT through the Centre de recherche en astrophysique du Qu\'ebec as well as the support from the Trottier Family Foundation and the Trottier Institute for Research on Exoplanets.\\
This project has received funding from the European Research Council (ERC) under the European Union's Horizon 2020 research and innovation programme (project {\sc Spice Dune}, grant agreement No 947634). This material reflects only the authors' views and the Commission is not liable for any use that may be made of the information contained therein.\\
This work has been carried out within the framework of the NCCR PlanetS supported by the Swiss National Science Foundation under grants 51NF40\_182901 and 51NF40\_205606.\\
FBa, \'EA, RD, LMa, TA, J-SM, JS-A \& PV acknowledges support from Canada Foundation for Innovation (CFI) program, the Universit\'e de Montr\'eal and Universit\'e Laval, the Canada Economic Development (CED) program and the Ministere of Economy, Innovation and Energy (MEIE).\\
ACar, XB, XDe, TF \& VY acknowledge funding from the French ANR under contract number ANR\-18\-CE31\-0019 (SPlaSH), and the French National Research Agency in the framework of the Investissements d'Avenir program (ANR-15-IDEX-02), through the funding of the ``Origin of Life" project of the Grenoble-Alpes University.\\
HC \& ML acknowledge support of the Swiss National Science Foundation under grant number PCEFP2\_194576.\\
ED-M, SCB, NCS, ARCS, EC \& JGd acknowledge the support from FCT - Funda\c{c}\~ao para a Ci\^encia e a Tecnologia through national funds by these grants: UIDB/04434/2020, UIDP/04434/2020.\\
ED-M further acknowledges the support from FCT through Stimulus FCT contract 2021.01294.CEECIND. ED-M acknowledges the support by the Ram\'on y Cajal grant RyC2022-035854-I funded by MICIU/AEI/10.13039/50110001103 and by ESF+.\\
ED-M, JGdS, NCS further acknowledge the support from FCT through the project 2022.04416.PTDC.\\
SCB acknowledges the support from Funda\c{c}\~ao para a Ci\^encia e Tecnologia (FCT) in the form of a work contract through the Scientific Employment Incentive program with reference 2023.06687.CEECIND.\\
The Board of Observational and Instrumental Astronomy (NAOS) at the Federal University of Rio Grande do Norte's research activities are supported by continuous grants from the Brazilian funding agency CNPq. This study was partially funded by the Coordena\c{c}\~ao de Aperfei\c{c}oamento de Pessoal de N\'ivel Superior—Brasil (CAPES) — Finance Code 001 and the CAPES-Print program.\\
BLCM \& AMM acknowledge CAPES postdoctoral fellowships.\\
BLCM acknowledges CNPq research fellowships (Grant No. 305804/2022-7).\\
NBC acknowledges support from an NSERC Discovery Grant, a Canada Research Chair, and an Arthur B. McDonald Fellowship, and thanks the Trottier Space Institute for its financial support and dynamic intellectual environment.\\
XDu acknowledges the support from the European Research Council (ERC) under the European Union’s Horizon 2020 research and innovation programme (grant agreement SCORE No 851555) and from the Swiss National Science Foundation under the grant SPECTRE (No 200021\_215200).\\
DE acknowledge support from the Swiss National Science Foundation for project 200021\_200726. The authors acknowledge the financial support of the SNSF.\\
JIGH, RR, ASM, FGT, NN, JLR \& AKS acknowledge financial support from the Spanish Ministry of Science, Innovation and Universities (MICIU) project PID2020-117493GB-I00.\\
ICL acknowledges CNPq research fellowships (Grant No. 313103/2022-4).\\
CMo acknowledges the funding from the Swiss National Science Foundation under grant 200021\_204847 “PlanetsInTime”.\\
JRM acknowledges CNPq research fellowships (Grant No. 308928/2019-9).\\
Co-funded by the European Union (ERC, FIERCE, 101052347). Views and opinions expressed are however those of the author(s) only and do not necessarily reflect those of the European Union or the European Research Council. Neither the European Union nor the granting authority can be held responsible for them.\\
JLA, RLG, DOF, YSM \& MAT acknowledge CAPES graduate fellowships.\\
We acknowledge funding from the European Research Council under the ERC Grant Agreement n. 337591-ExTrA.\\
ARCS acknowledges the support from Funda\c{c}ao para a Ci\^encia e a Tecnologia (FCT) through the fellowship 2021.07856.BD.\\
LD acknowledges the support of the Natural Sciences and Engineering Research Council of Canada (NSERC) [funding reference number 521489] and from the Fonds de recherche du Qu\'ebec (FRQ) - Secteur Nature et technologies [funding file number 332355].\\
DBF acknowledges financial support from the Brazilian agency CNPq-PQ (Grant No. 305566/2021-0). Continuous grants from the Brazilian agency CNPq support the STELLAR TEAM of the Federal University of Ceara's research activities.\\
FG acknowledges support from the Fonds de recherche du Qu\'ebec (FRQ) - Secteur Nature et technologies under file \#350366.\\
AL acknowledges support from the Fonds de recherche du Qu\'ebec (FRQ) - Secteur Nature et technologies under file \#349961.\\
LMo acknowledges the support of the Natural Sciences and Engineering Research Council of Canada (NSERC), [funding reference number 589653].\\
NN acknowledges financial support by Light Bridges S.L, Las Palmas de Gran Canaria.\\
NN acknowledges funding from Light Bridges for the Doctoral Thesis "Habitable Earth-like planets with ESPRESSO and NIRPS", in cooperation with the Instituto de Astrof\'isica de Canarias, and the use of Indefeasible Computer Rights (ICR) being commissioned at the ASTRO POC project in the Island of Tenerife, Canary Islands (Spain). The ICR-ASTRONOMY used for his research was provided by Light Bridges in cooperation with Hewlett Packard Enterprise (HPE).\\
CPi acknowledges support from the NSERC Vanier scholarship, and the Trottier Family Foundation. This work was funded by the Institut Trottier de Recherche sur les Exoplan\`etes (iREx).\\
AP acknowledges support from the Unidad de Excelencia Mar\'ia de Maeztu CEX2020-001058-M programme and from the Generalitat de Catalunya/CERCA.\\
JS-F acknowldges financial support from the Agencia Estatal de Investigaciòn (AEI/10.13039/501100011033) of the Ministerio de Ciencia e Innovaci\'on and the ERDF "A way of making Europe" through project PID2022-137241NB-C42.\\
AKS acknowledges financial support from La Caixa Foundation (ID 100010434) under the grant LCF/BQ/DI23/11990071.\\
TV acknowledges support from the Fonds de recherche du Qu\'ebec (FRQ) - Secteur Nature et technologies under file  \#320056.

\end{acknowledgements}


\bibliographystyle{aa}
\bibliography{NIRPS_WASP69} 

\begin{appendix}

\section{Photometric observations}
\begin{figure}[h!]
\includegraphics[width=\columnwidth]{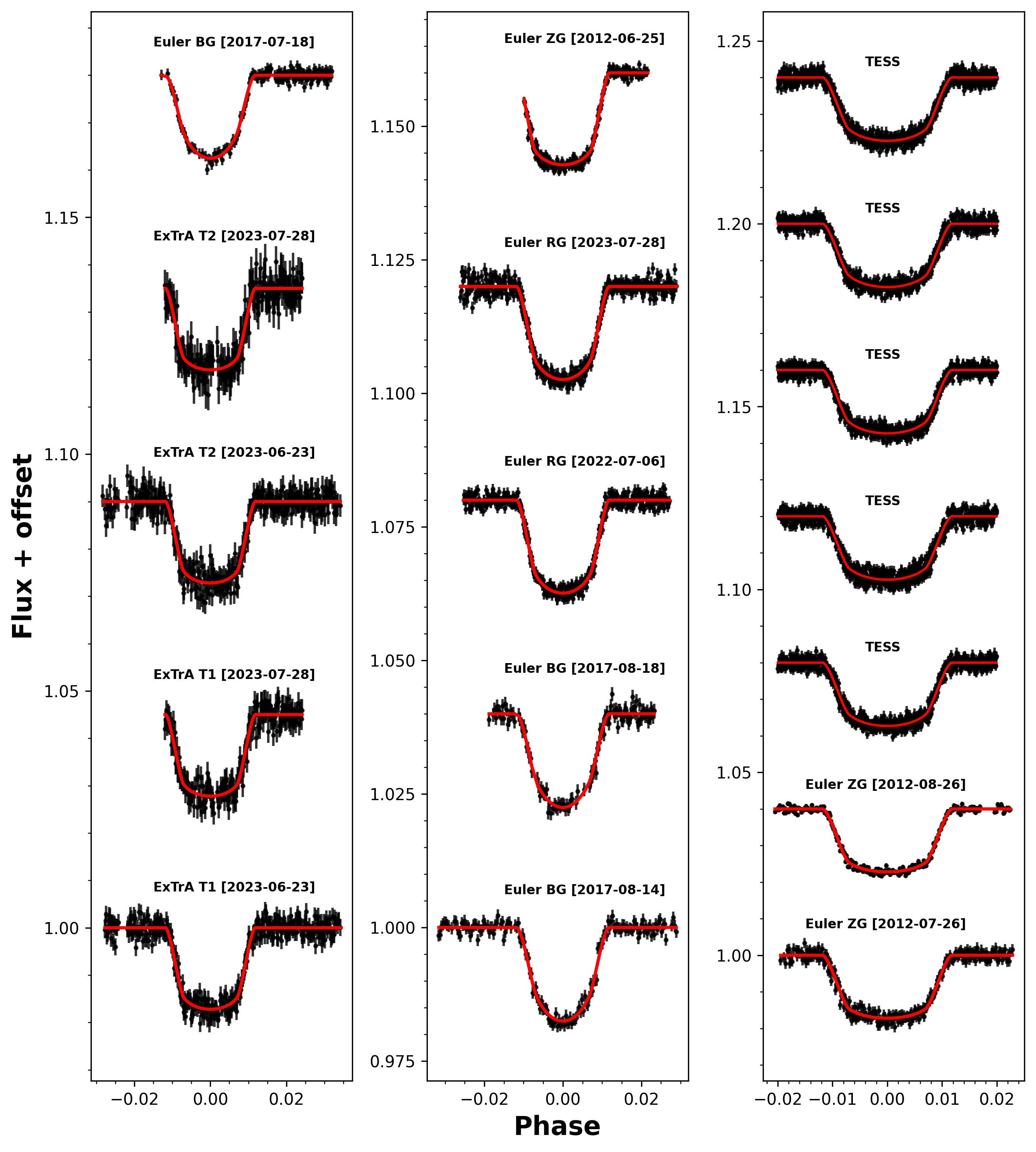}
\centering
\caption[]{De-trended light curves obtained from EulerCam, ExTrA, and TESS. The best-fit transit models are shown in red.}
\label{fig:photo_models}
\end{figure}

Figure\,\ref{fig:photo_models} displays all the light curves used in this analysis to refine the parameters and detect the presence of spot-crossing events. Table\,\ref{tab:phot_prior} shows the priors used to model the light curves.
\begin{table}[h!]
    \centering
    \caption{Priors on stellar and planetary parameters}
    \begin{tabular}{ccc}
        \toprule \toprule
        Parameter & Units & Prior type and range\\
        \toprule \toprule
         Stellar density ($\rho{_\star}$) & g cm$^{-3}$ & 
         $\mathcal{U}$(1, 6) \\
         Planet-to-star ratio (R$_{p}$/R$_{\star}$)  &     & $\mathcal{U}$(0, 0.2) \\
         Impact parameter (b) & & $\mathcal{U}$(0, 1) \\
         Mid-transit time ($T_{0}$) & BJD$_{\rm{TDB}}$ & $\mathcal{N}$(2455748.83, 0.05) \\
         Orbital period (P$_{\rm{orb}}$) & days & $\mathcal{N}$(3.87, 0.01)\\

         Stellar mass (M$_{\star}$) & M$_{\odot}$ &  $\mathcal{N}$(0.831, 0.02) \\

         q1$_{\rm{ExTrA}}$ & & $\mathcal{N}$(0.31, 0.02) \\
         q2$_{\rm{ExTrA}}$ & & $\mathcal{N}$(0.28, 0.04) \\
         q1$_{\rm{Euler_B}}$ & & $\mathcal{N}$(0.75, 0.01) \\
         q2$_{\rm{Euler_B}}$ & & $\mathcal{N}$(0.57, 0.01) \\
         q1$_{\rm{Euler_R}}$ & & $\mathcal{N}$(0.52, 0.02) \\
         q2$_{\rm{Euler_R}}$ & & $\mathcal{N}$(0.40, 0.03) \\
         q1$_{\rm{Euler_Z}}$ & & $\mathcal{N}$(0.31, 0.02) \\
         q2$_{\rm{Euler_Z}}$ & & $\mathcal{N}$(0.31, 0.03) \\
         q1$_{\rm{TESS}}$ & & $\mathcal{N}$(0.43, 0.02) \\
         q2$_{\rm{TESS}}$ & & $\mathcal{N}$(0.36, 0.03) \\

         Spot longitude & $^{\circ}$ & $\mathcal{U}$(-180, 180) \\
         Spot latitude  & $^{\circ}$ & $\mathcal{U}$(-90, 90) \\
         Spot size      & $^{\circ}$ & $\mathcal{U}$(0, 20) \\
         Spot contrast  &            & $\mathcal{U}$(0, 1) \\
        \toprule
        \end{tabular}
    \label{tab:phot_prior}
\end{table}

\section{Rossiter McLaughlin effect}

\begin{figure*}[h!]
\includegraphics[width=0.7\textwidth]{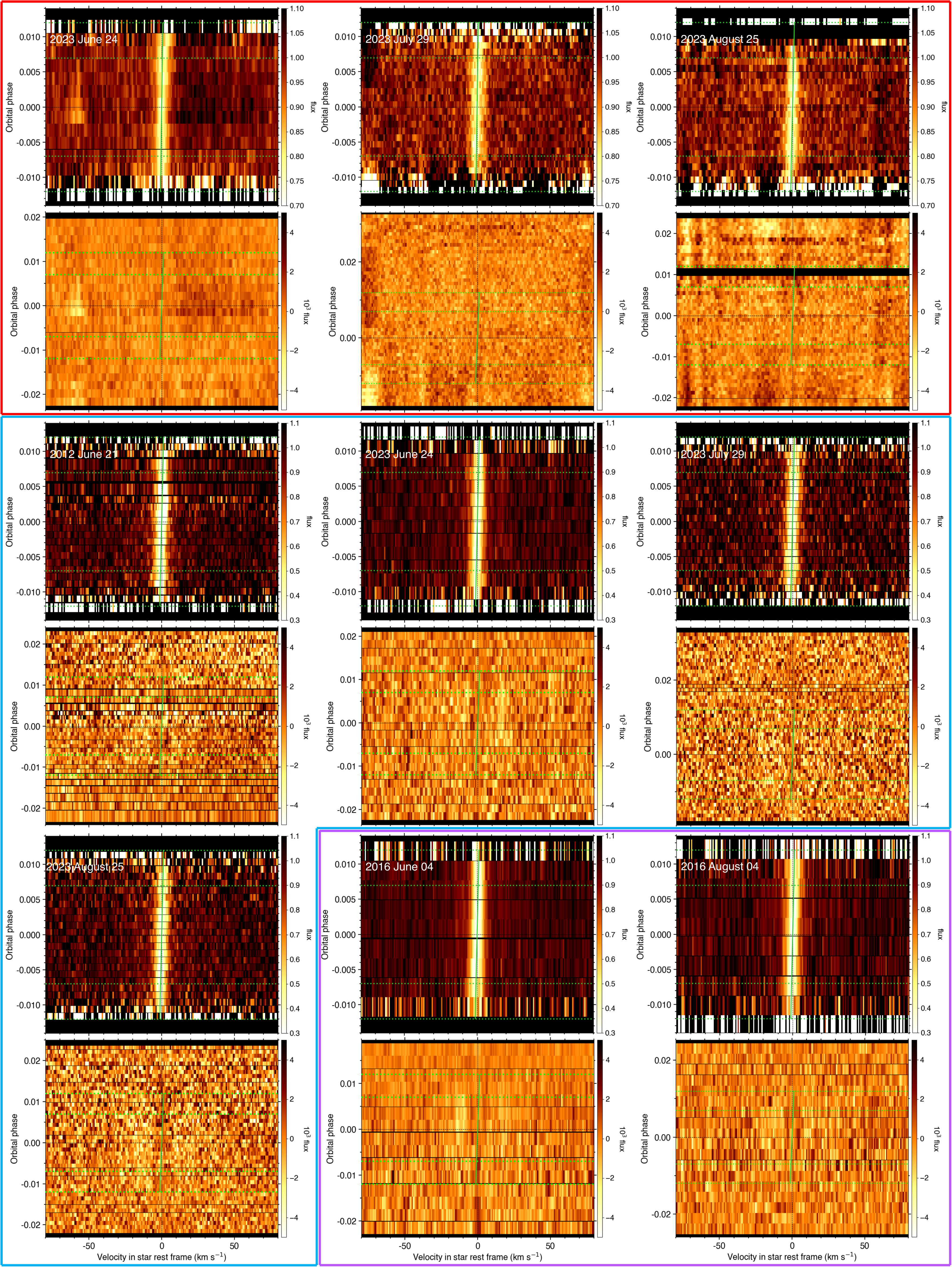}
\centering
\caption[]{Maps of WASP-69 intrinsic CCF profiles measured in individual epochs with NIRPS (red frame), HARPS (blue frame), and HARPS-N (purple frame), plotted as a function of RV in the star rest frame (abscissa) and orbital phase (ordinate). Each map is associated with a residual map showing the difference between out-of-transit CCFs and their master or between intrinsic CCFs and their best-fit model (scaled to the level of out-of-transit residuals). Horizontal dashed green lines show transit contacts. The solid green line indicates the surface RV track associated with the best RMR fit. }
\label{fig:Intr_maps}
\end{figure*}

Table\,\ref{table:detrending} shows for each visit how the stellar line properties have been detrended. Figure\,\ref{fig:Intr_maps} presents the intrinsic CCF profile maps of WASP-69 for each transit observed with NIRPS, HARPS, and HARPS-N.
\onecolumn
\begin{table*}[h!]
\begin{minipage}[h!]{\textwidth}
\caption{Detrending of the WASP-69 stellar line}
\label{table:detrending}
\small
\centering
\begin{tabular}{l c c c | c c c c | c c }
\toprule \toprule
Instrument    &  \multicolumn{3}{c}{NIRPS}  &  \multicolumn{4}{c}{HARPS}  &  \multicolumn{2}{c}{HARPS-N} \\
Visit         &  1      & 2     & 3         &           1 & 2 & 3  & 4    &   1 & 2  \\
Night & 2023-06-23  & 2023-07-28 & 2023-08-24 & 2012-06-21 & 2023-06-23  & 2023-07-28 & 2023-08-24 & 2016-06-03 & 2016-08-04\\
\toprule \toprule
Contrast      &  -      & -     & Phase     &   Phase   & S/N   & S/N   & S/N   &   S/N & -  \\
FWHM          &  Phase  & Phase & Phase     &   -       & Phase & -     & Phase &   S/N & -  \\
RV            &  Phase  & Phase & -         &  Phase    & -     & -     & -     &   -   & -  \\
\bottomrule
\end{tabular}
\tablefoot{This table indicates whether the out-of-transit series of a given property (contrast, FWHM, and RV centroid) describing the Gaussian fit to disk-integrated CCFs correlates linearly with phase or S/N or remains stable. 
}
\end{minipage}
\end{table*}

\twocolumn
\section{Comparison between the two pipelines} \label{app:pipelines_comp}
\begin{figure}[h!]
\includegraphics[width=\columnwidth]{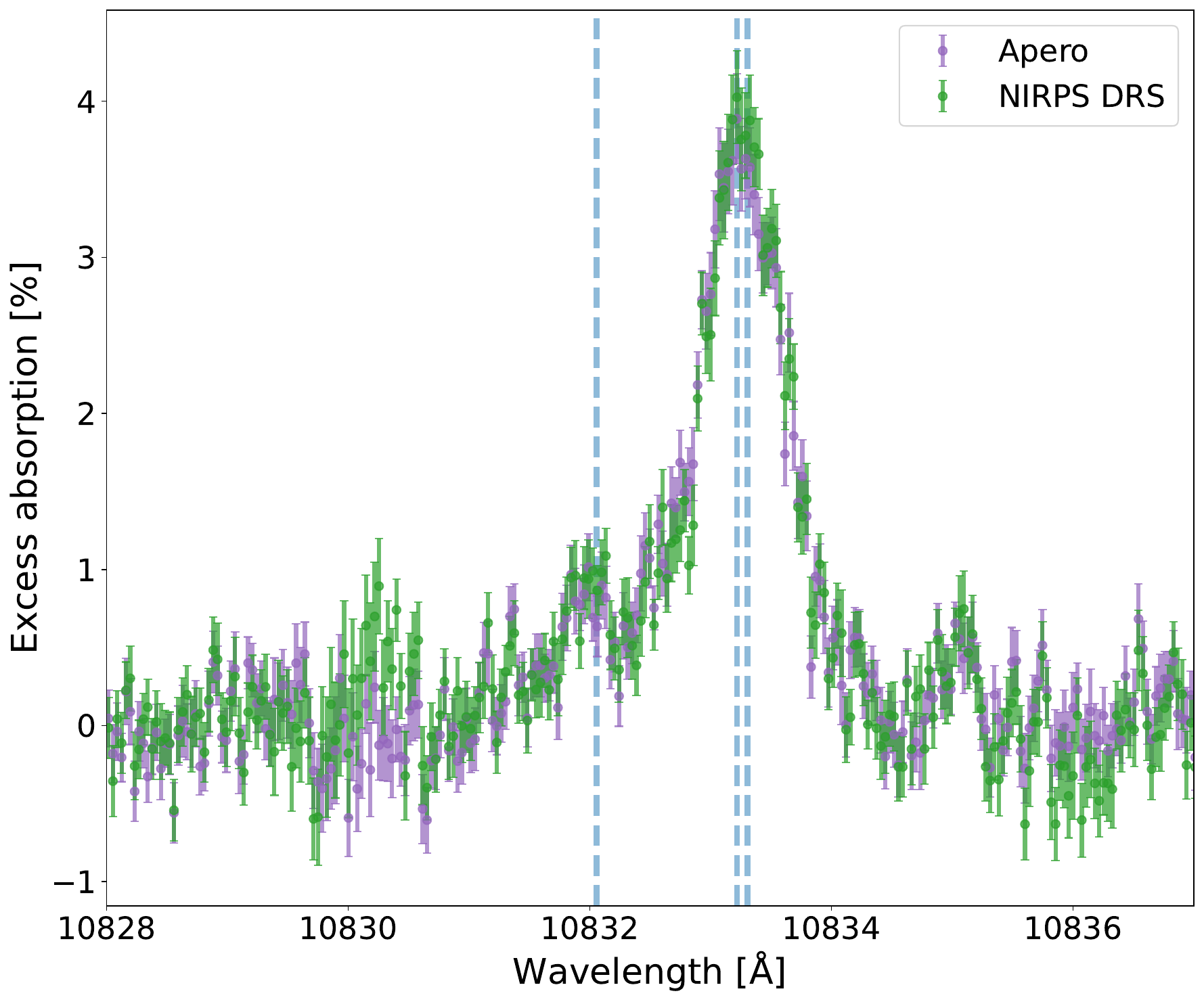}
\centering
\caption[]{Comparison of the transmission spectrum obtained with the APERO (green) and NIRPS DRS (purple) pipelines. The vertical blue dashed lines indicate the helium lines' positions.}
\label{fig:TS_pipelines_comparison}
\end{figure}

Figure\,\ref{fig:TS_pipelines_comparison} presents the transmission spectrum of WASP-69\,b around the helium triplet, where the spectra are extracted either with the APERO pipeline \citep{cook_apero_2022} or the NIRPS DRS pipeline \citep{pepe_espresso_2021}. The results are in good agreement.

\section{Computation of the helium signature intensity}\label{app:excess_computation}
In this appendix, we provide the readers with Table\,\ref{tab:tech_abs}, which summarizes different techniques used in the literature to assess the helium triplet signature strength. It includes the maximum absorption of the signature, the integrated flux over passbands of 0.75\,\AA\ and 1\,\AA\ centered at 10833.22\,\AA , a Gaussian fit, and the equivalent width computed from 10828 to 10837\,\AA. We also want to highlight that the equivalent width remains sensible to the spectral range selected and, thus, to the systematics and features within.

\begin{table}[h!]
    \centering
    \begin{tabular}{cc}
        \toprule \toprule
        Technique & Excess absorption\\
        \toprule \toprule
        Maximum absorption [\%] & 4.02$\pm$0.24\\
        Integrated 0.75\,\AA\ passband [\%] & 3.17$\pm$0.05\\
        Integrated 1.00\,\AA\ passband [\%] & 2.81$\pm$0.05\\
        Gaussian fit [\%] & 3.48$\pm$0.10\\
        Equivalent width [m\AA] & 49.18$\pm$1.22\\
        \toprule
        \end{tabular}
    \caption{Helium signature strength based on different techniques. The signal integrated over a 0.75\,\AA\ passband is the technique chosen in this paper.}
    \label{tab:tech_abs}
\end{table}

\section{Atmospheric escape modelling}\label{app:escape}

\begin{figure}[h!]
     \centering
     \begin{minipage}[t]{0.475\textwidth}
         \centering
         \includegraphics[width=\textwidth]{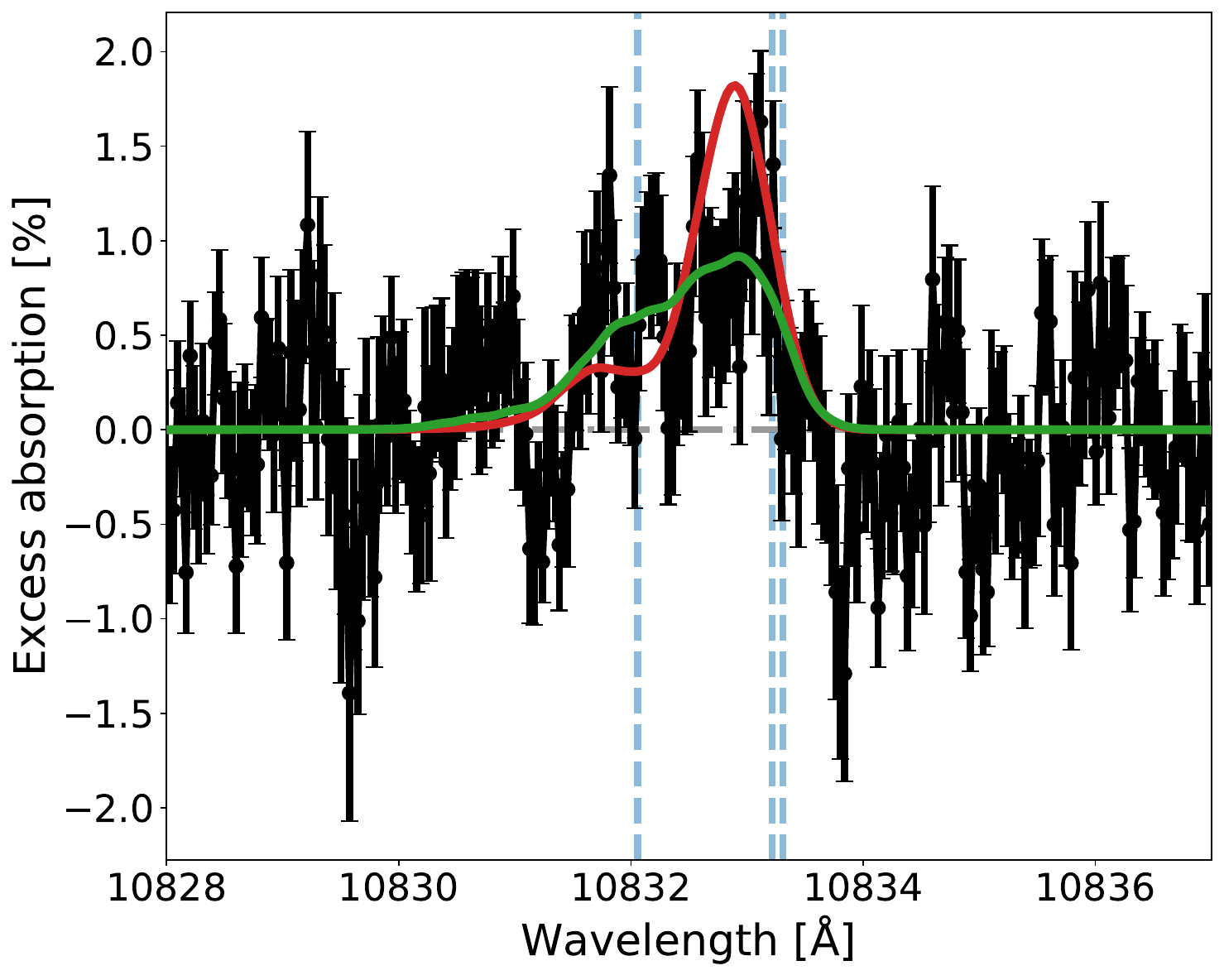}
     \end{minipage}
     \\
     \begin{minipage}[t]{0.52\textwidth}
         \centering
         \includegraphics[width=\textwidth]{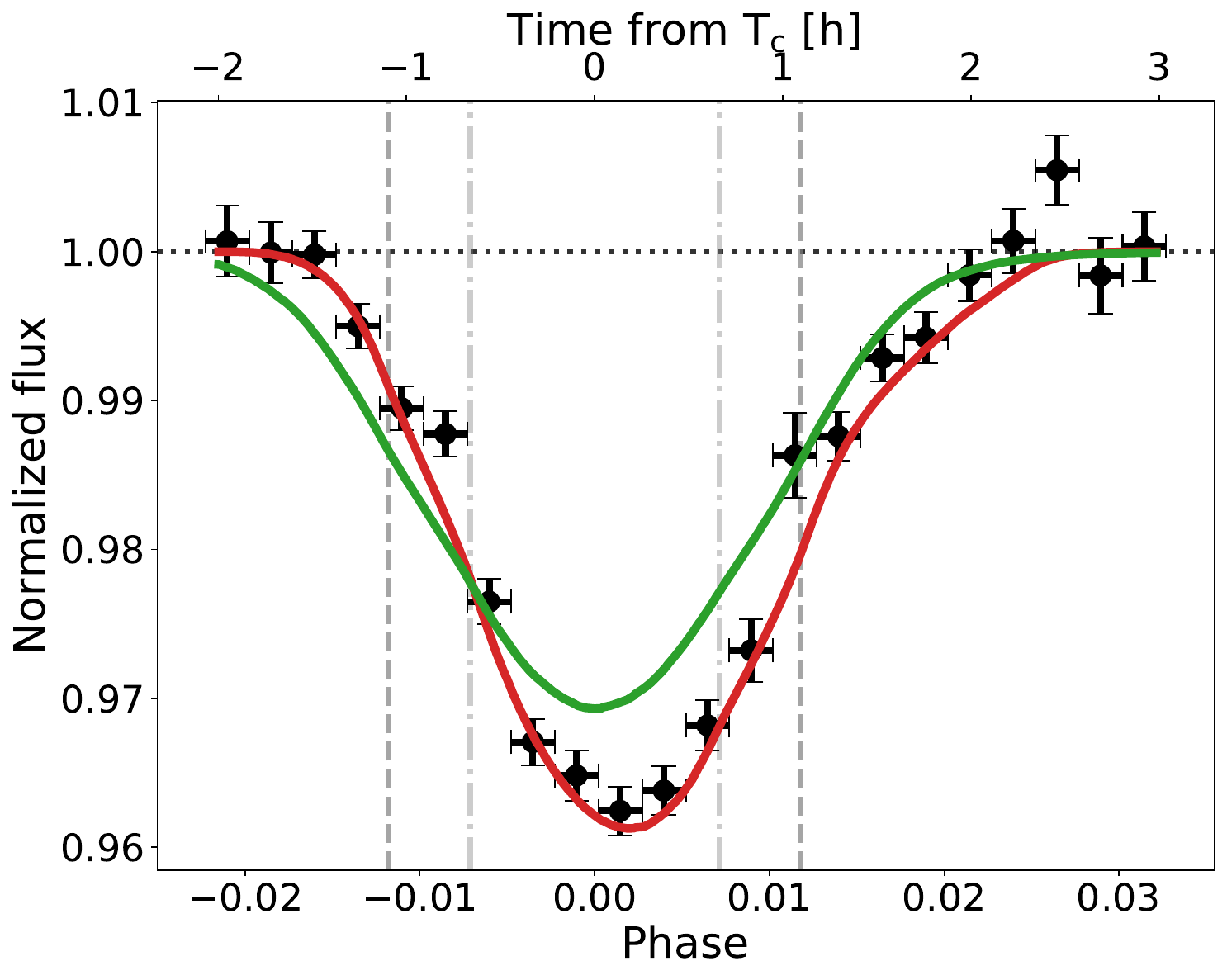}
     \end{minipage}
        \caption{Comparison of the post-transit absorption signal with EVE models including thermosphere and exosphere. In black are the data, in red is the best-fit model over the whole timeseries, and in green is the best-fit model adjusted to the post-transit from t$_4$ to 50 minutes after t$_4$. The \textit{top} panel shows the transmission spectrum and the \textit{bottom} panel shows the light curve.}
        \label{fig:bestfit_post_transit}
\end{figure}

To test the idea that the exosphere might be subject to repopulation within the cometary tail, we tried to set the metastable helium natural de-excitation time to zero. In this case, the metastable helium of the exosphere is depopulated only through photoionization. This appendix presents how the EVE models are able to reproduce the post-transit absorption as defined from t$_4$ to 50 minutes after t$_4$. We compare in Fig.\,\ref{fig:bestfit_post_transit}, the transmission spectrum built over this specific timeframe (left panel) and the helium light curve (right panel, also presented in Fig.\,\ref{fig:LC}) with the best-fit model estimated over the whole timeseries in red (also shown in Figs.\,\ref{fig:TS} and \ref{fig:LC}) and the best-fit model adjusted on the post-transit absorption signal in green. We see that the latest model is able to slightly better fit both the transmission spectrum and light curve signal post-transit. We note, however, that this model reproduces the full transit less accurately than simulations presented in Section~\ref{sec:helium}.

\onecolumn
\section{Impact of the telluric correction on the master-out spectra}
\begin{figure}[h!]
\includegraphics[width=\textwidth]{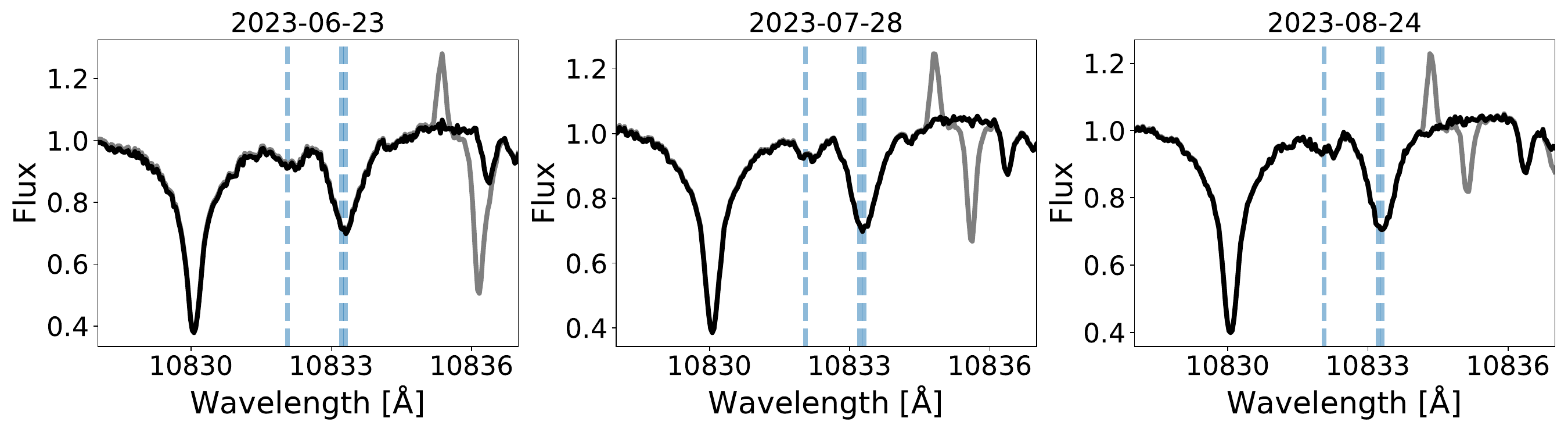}
\centering
\caption[]{Master-out spectrum for each night before (grey) and after (black) telluric correction for H$_2$O absorption and OH emission lines.}
\label{fig:master_tell}
\end{figure}

Figure\,\ref{fig:master_tell} presents the impact of the telluric correction on the master-out spectrum of each transit. Telluric lines are well disentangled from the stellar helium line, and their correction is consistent with the level of the stellar continuum.

\end{appendix}

\end{document}